\documentclass[sigconf,10pt]{acmart}
\usepackage{backnaur}
\usepackage[ruled,vlined]{algorithm2e}
\usepackage{graphicx}
\setcopyright{none}
\renewcommand\footnotetextcopyrightpermission[1]{}

\AtBeginDocument{%
  }

\setcopyright{acmlicensed}
\copyrightyear{2026}
\acmYear{2027}
\acmDOI{X.X}

\usepackage{graphicx}
\usepackage{pgfplots}
\pgfplotsset{compat=1.18}
\usepgfplotslibrary{colorbrewer}

\definecolor{oiBlack}{RGB}{0,0,0}
\definecolor{oiOrange}{RGB}{230,159,0}
\definecolor{oiSkyBlue}{RGB}{86,180,233}
\definecolor{oiBluishGreen}{RGB}{0,158,115}
\definecolor{oiYellow}{RGB}{240,228,66}
\definecolor{oiBlue}{RGB}{0,114,178}
\definecolor{oiVermillion}{RGB}{213,94,0}
\definecolor{oiReddishPurple}{RGB}{204,121,167}

\pgfplotsset{
    asplos_style/.style={
        width=\columnwidth,
        height=0.7\columnwidth,
        legend style={at={(0.5,-0.25)}, anchor=north, legend columns=-1, font=\footnotesize},
        label style={font=\small},
        tick label style={font=\footnotesize},
        grid=major,
        grid style={dashed, gray!30},
        axis lines*=left, 
        cycle list={ 
            {oiBlue, mark=*, thick},
            {oiOrange, mark=square*, thick},
            {oiGreen, mark=triangle*, thick}
        }
    }
}
\usepackage{listings}
\usepackage{xcolor}

\usepackage{booktabs} 
\usepackage{pgfplotstable} 
\pgfplotsset{compat=1.18}
\usepackage{float}

\pgfplotstableset{
    asplos_table/.style={
        col sep=comma, 
        string type,    
        header=true,
        every head row/.style={before row=\toprule, after row=\midrule},
        every last row/.style={after row=\bottomrule},
        column type=l,  
    }
}
\usepackage{subcaption}

\begin{document}


\newcommand{\graphdd}{%
  \begin{tikzpicture}[
      vertex/.style={
        rectangle,
        draw,
        minimum width=0.25cm,
        minimum height=0.5cm,
        font=\footnotesize,
        fill=green!50,
        text=black,
    },
      edg/.style={->, thick, black},
      border/.style={rectangle, draw=black, minimum height=2.8cm, minimum width=3cm}
  ]
  \node[border] (src) at (0,0) {};
  \node[vertex] (graph) at (-1,0) {graph};
  \node[vertex] (rma) at (0.25,1) {rma};
  \node[vertex] (gp) at (0.25,0.2) {gp};
  \node[vertex] (np) at (-0.125,-0.8) {np};
  \node[vertex] (ep) at (0.8,-0.8) {ep};
  \draw[edg] (rma)-- (gp);
  \draw[edg] (gp) -- (np);
  \draw[edg] (gp) -- (ep);
  \draw[edg] (gp) -- (graph);
  \end{tikzpicture}%
}

\def\xshift{-2cm}
\def\yshift{-3cm}

\newcommand{\graphndd}{%
  \begin{tikzpicture}[
      vertex/.style={
        rectangle,
        draw,
        minimum width=0.5cm,
        minimum height=0.25cm,
        font=\footnotesize,
        fill=green!50,
        text=black,
    },
      edg/.style={->, thick, black},
      border/.style={rectangle, draw=black, minimum height=3cm, minimum width=4.2cm}
  ]
  \node[border] (src) at (-0.7, 0) {};
  \node[vertex] (graph) at (0,-1) {graph};
  \node[vertex] (rma) at (1,0.25) {rma};
  \node[vertex] (gp) at (0.2,0.25) {gp};
  \node[vertex] (np) at (-0.8,-0.125) {np};
  \node[vertex] (ep) at (-0.8,0.8) {ep};
  \node[vertex, fill=blue!30] (blkR) at (-1.8, -0.2) {blkR};
  \node[vertex, fill=red!40, rotate =90] (gs) at (-2.5, 0) {Global struct singleton};
  \draw[edg] (rma)-- (gp);
  \draw[edg] (gp) -- (np);
  \draw[edg] (gp) -- (ep);
  \draw[edg] (gp) -- (graph);
  \draw[edg] (blkR) -- (graph);
  \draw[edg] (blkR) -- (np);
  \draw[edg] (blkR) -- (ep);
  \end{tikzpicture}%
}
\newcommand{\graphds}{
\begin{tikzpicture}[
  container/.style={rectangle, draw=orange!80!black, fill=orange!20, minimum width=1cm, minimum height=1.33cm},
  circlestyle/.style={circle, draw, minimum size=0.75cm, fill=yellow!20},
  box/.style={rectangle, draw, minimum width=0.1cm, minimum height=0.2cm, fill=white},
  edge/.style={->, thick, black}
]
  \node[container] (c1) at (1.5*1, 0) {};
  \node[container] (c2) at (1.5*2, 0) {};
  \node[container] (c3) at (1.5*3, 0) {};
  \node[circlestyle] (circle1) at ([yshift=-0.45cm]c1.north) {\small{sg1}};
  \node[circlestyle] (circle2) at ([yshift=-0.45cm]c2.north) {\small{sg2}};
  \node[circlestyle] (circle3) at ([yshift=-0.45cm]c3.north) {\small{sg3}};
  \node[box] at ([yshift=0.25cm,xshift=0.3cm*-1]c1.south) (A1) {};
  \node[box] at ([yshift=0.25cm,xshift=0.3cm*0]c1.south) (A2){};
  \node[box] at ([yshift=0.25cm,xshift=0.3cm*1]c1.south) (A3){};
  \node[box] at ([yshift=0.25cm,xshift=0.3cm*-1]c2.south) (A4){};
  \node[box] at ([yshift=0.25cm,xshift=0.3cm*0]c2.south) (A5){};
  \node[box] at ([yshift=0.25cm,xshift=0.3cm*1]c2.south) (A6){};
  \node[box] at ([yshift=0.25cm,xshift=0.3cm*-1]c3.south) (A7){};
  \node[box] at ([yshift=0.25cm,xshift=0.3cm*0]c3.south) (A8){};
  \node[box] at ([yshift=0.25cm,xshift=0.3cm*1]c3.south) (A9){};
  \draw[edge] (circle1) -- (A4);
  \draw[edge] (circle1) -- (A5);
  \draw[edge] (circle1) -- (A9);
  \draw[edge] (circle2) -- (A1);
  \draw[edge] (circle2) -- (A8);
  \draw[edge] (circle2) -- (A7);
  \draw[edge] (circle3) -- (A5);
  \draw[edge] (circle3) -- (A8);
\end{tikzpicture}
}

\newcommand{\astdia}{
\begin{tikzpicture}[
    vertex/.style={
        rectangle,
        draw,
        minimum width=0.25cm,
        minimum height=0.5cm,
        font=\footnotesize,
        fill=green!10,
        text=black,
    },
    edge/.style={->,thick,blue},
    dots/.style={font=\footnotesize}
]

\node[vertex] (func) at (4,0) {func};
\node[vertex] at(2.5,-0.7) (func_sig) {sig};
\node[vertex] at(2.5,-2.2) (sig) {sig};
\node[vertex] at(4.5,-0.7) (func_bod) {blk};
\draw[edge] (func) -- (func_sig);
\draw[edge] (func) -- (func_sig);
\draw[edge] (func) -- (func_bod);
\node[vertex] at(2.3,-1.4) (while) {whl};
\node[vertex] at(3.2,-1.4) (for) {for};
\node[vertex] at(4,-1.4) (if) {if};
\node[vertex] at(4,-2.2) (cond) {cond};
\node[vertex] at(3.5,-2.8) (body) {body};
\node[vertex] at(5,-1.4) (decl) {decl};
\node[vertex] at(6,-1.4) (asgn) {asgn};
\node[vertex] at (5.5, -2.2) (expr) {expr};
\draw[edge] (func_bod) -- (while);
\draw[edge] (func_bod) -- (for);
\draw[edge] (func_bod) -- (if);
\draw[edge] (func_bod) -- (decl);
\draw[edge] (func_bod) -- (asgn);
\draw[edge] (asgn) -- (expr);
\draw[edge]  (decl) -- (expr);
\draw[edge]  (if) -- (cond);
\draw[edge]  (if) -- (body);
\draw[edge]  (body) to[out=360, in=360] (func_bod);
\draw[edge] (for) -- (sig);
\draw[edge] (while) -- (sig);
\draw[edge] (while) -- (body);
\draw[edge] (for) -- (body);

\end{tikzpicture}
}
\newcommand{\atoa}{
    \begin{tikzpicture}[
            proc/.style={
            rectangle,
            draw,
            minimum width=0.45cm,
            minimum height=0.5cm,
            font=\footnotesize,
            fill=green!30,
            text=black,
            rounded corners
            },
            comm/.style={->,thick,black},
            dots/.style={font=\footnotesize}
        ]
        \node[proc] (A1) at (0,0) {P0};
        \node[proc, right=0.4cm of A1] (A2) {P1};
        \node[dots, right=0.05cm of A2] (A4) {...};
        \node[proc, right=0.05cm of A4] (AN) {PN};
        
        \node[proc, below=0.6cm of A1] (B1) {P0};
        \node[proc, below=0.6cm of A2] (B2) {P1};
        \node[dots, below=0.6cm of A2] (B4) {...};
        \node[proc, below=0.6cm of AN] (BN) {PN};
        
        \draw[comm] (A1) -- (B2);
        \draw[comm] (A1) -- (BN);
        
        \draw[comm] (A2) -- (B1);
        \draw[comm] (A2) -- (BN);
        
        
        \draw[comm] (AN) -- (B1);
        \draw[comm] (AN) -- (B2);
        \end{tikzpicture}%
}
\newcommand{\rmasync}{
\begin{tikzpicture}[
        proc/.style={
            rectangle,
            draw,
            minimum width=0.45cm,
            minimum height=0.5cm,
            font=\footnotesize,
            fill=green!30,
            text=black,
            rounded corners
        },
        comm/.style={->,thick,black},
        dots/.style={font=\footnotesize}
    ]
    \node[proc] (A1) at (0,0) {P0};
    \node[proc, right=0.4cm of A1] (A2) {P1};
    \node[dots, right=0.05cm of A2] (A4) {...};
    \node[proc, right=0.05cm of A4] (AN) {PN};
    
    \node[proc, below=0.6cm of A1] (B1) {P0};
    \node[proc, below=0.6cm of A2] (B2) {P1};
    \node[dots, below=0.6cm of A4] (B4) {...};
    \node[proc, below=0.6cm of AN] (BN) {PN};
    
    \draw[comm] (A1) -- (BN);
    \draw[comm] (A2) -- (BN);
    \draw[comm] (AN) -- (B1);
    \end{tikzpicture}
}
\newcommand{\motivationInterSSSP}{
\begin{figure}
    \label{graph:commpat}
    \begin{tikzpicture}
        \begin{axis}
            
        \end{axis}
    \end{tikzpicture}
\end{figure}
}

\newcommand{\experimentScaleRedNat}{
\begin{figure}
    \label{graph:srednat}
    \begin{tikzpicture}
        \begin{axis}
        \end{axis}
    \end{tikzpicture}
\end{figure}
}
\newcommand{\experimentScaleRedSyn}{
\begin{figure}
    \begin{tikzpicture}
        \begin{axis}
            
        \end{axis}
    \end{tikzpicture}
\end{figure}
}

\newcommand{\experimentCompSSSP}{
\begin{figure}
    \begin{tikzpicture}
        \begin{axis}
            
        \end{axis}
    \end{tikzpicture}
\end{figure}
}

\newcommand{\experimentCompWCC}{
\begin{figure}
    \begin{tikzpicture}
        \begin{axis}
            
        \end{axis}
    \end{tikzpicture}
\end{figure}
}

\newcommand{\experimentCompPR}{
\begin{figure}
    \begin{tikzpicture}
        \begin{axis}
            
        \end{axis}
    \end{tikzpicture}
\end{figure}
}


\title{StarDist: A Code Generator for Distributed Graph
Algorithms}

\author{Barenya Kumar Nandy}
\email{bknandy98@gmail.com}
\orcid{1234-5678-9012}
\affiliation{%
  \institution{Indian Institute of Technology}
  \city{Madras}
  \state{Chennai}
  \country{India}
}

\author{Rupesh Nasre}
\orcid{1234-5678-9012}
\email{rupesh@cse.iitm.ac.in}
\affiliation{%
  \institution{Indian Institute of Technology}
  \city{Madras}
  \state{Chennai}
  \country{India}
}


\begin{abstract}
  We introduce StarDist, a Domain Specific Language for generating high-performant distributed graph algorithms in the message passing model. Our analysis-transformation framework optimizes graph traversal based on graph property access patterns, reduces global lock acquisitions on distributed structures, and minimizes message queues used in reduction operations. We provide a network optimized communication runtime for reduction operations that couples with our analysis framework and optimizes the propagation of updates based on vertex residency. StarDist is able to identify monotonic reduction blocks and is able to fuse reduction iterations over graphs into \textit{pulses}. We evaluate StarDist 
  using three fundamental graph algorithms belonging to the CONGEST model: single-source shortest paths, weakly connected components, and PageRank computation, using a suite comprising both real-world and synthetic graphs across varying densities of topological compaction. Our results illustrate that the code generated with StarDist outperforms the distributed frameworks DRONE and D-Galois by an average of 19$\times$ and 7$\times$, respectively on our high communication setup and by 1.4$\times$ and 1.92$\times$ respectively on our high congestion network setup when averaged across all three algorithms.

\end{abstract}



\received{15 April 2026}

\maketitle

\section{Introduction}
 Semi-structured and unstructured data, occurring routinely in the real world, often benefit from being cast as graphs, which provide logical abstractions that make analytical derivations simpler. As graphs get larger, the \textit{irregular} memory access patterns exhibited by most graph algorithms result in poor spatial and temporal localities. This results in frequent instruction stalls due to cache misses~\cite{ref9}. The vertex centric view of shared memory graph algorithms~\cite{Greenmarl, Galois, Galois2, Ligra} for large graphs would lead to a large number of threads on existing cores and levy high synchronization costs. When shared memory graph algorithms operate on graphs that exhaust main memory, they are either terminated by the Operating System~\cite{ref26} or they experience massive slowdowns due to demand paging~\cite{scalegraph}. Data centers are usually configured with storage area networks with considerably low IO bandwidths compared to the high bandwidth that can be achieved through RDMA combined with remote RAMs~\cite{ref24}, indicating a trend that network IO between resident memory regions is faster than page faults on disks. While out-of-core solutions have been proposed~\cite{ref22}, distributed processing of graph algorithms has become a widely accepted solution for mitigating physical memory limitations~\cite{survey}.

Despite the necessity of distributed execution, programming these systems from scratch remains a  challenge. The Message Passing Interface standard~\cite{ref7} provides communication directives for a uniform abstraction across different communication stacks. Unlike traditional distributed workloads that exhibit predictable communication patterns, graph algorithms generate highly irregular, any-to-any traffic that fluctuates dynamically~\cite{ref4, powergraph}. This irregularity is driven by the fact that most \textit{natural} graphs—derived from social networks, web crawls, and biological data exhibit a power-law degree distribution, where a small fraction of \textit{hub} vertices is connected to a vast majority of the graph~\cite{Faloutsos1999}. These hubs create pathological network hotspots, transacting a high volume of small updates, converging on a small set of vertices simultaneously. This floods the bandwidth and result in network latencies. While high-level frameworks such as Pregel~\cite{Pregel}, PowerGraph~\cite{powergraph}, D-Galois~\cite{ref1} and DRONE~\cite{DRONE} attempt to hide programming complexity, they fail to exploit the zero-copy capabilities of modern one-sided Remote Memory Access (RMA) primitives. To ensure correctness, graph compilers such as 
StarPlat \cite{ref5, ref6} and Fregel~\cite{fregel} generate distributed programs that rely on pessimistic MPI-RMA locking protocols or point-to-point directives to serialize access.

In this paper, we present StarDist, a distributed graph compiler backend that uses static analysis to resolve this fundamental tension and tame network irregularity. StarDist analyzes the Abstract Syntax Tree (AST) to identify reduction operations that allow decoupling local execution from network-level synchronization. By recognizing these semantics, StarDist safely bypasses pessimistic MPI-RMA locks for intra-partition updates, applying them via direct, zero-overhead memory accesses. For inter-partition communication, StarDist mitigates incast events by dynamically reducing and aggregating outbound network traffic. By optimizing message queue sizes for remote updates, StarDist prevents pathological hotspots that traditionally overwhelm network interface cards (NICs). StarDist delivers the familiarity of a high-level programming language—employing programming directives native to the Boost Graph Library (BGL)~\cite{ref31} while generating highly optimized distributed code optimized.
To this end, this paper makes the following contributions:
\begin{enumerate}
    \item A distributed runtime utilizing passive MPI-RMA windows, a distributed worklist, and dynamic message aggregation for outbound traffic, preventing NIC hotspots.
    \item A static analysis pattern that identifies AST reduction semantics to bypass pessimistic synchronization, enabling zero-overhead direct memory accesses for intra-partition reads and writes.
    \item A compiler transformation pass that fuses multiple reduction iterations into \textit{pulses}.
    \item A topology-aware analysis framework that dynamically swaps between vertex and edge-centric iterations to minimize wasted cpu cycles during graph scans.
\end{enumerate}

To test our distributed code generation, we focus on three core algorithms bounded by the CONGEST model~\cite{congest}, in particular: Single-Source Shortest Path (SSSP), Weakly Connected Components (WCC), and PageRank (PR). We benchmark StarDist against state-of-the-art frameworks, namely D-Galois and DRONE, across two distinct hardware topologies to evaluate both scale-up and scale-out efficiency: a compute-dense configuration (3 nodes with 32-core Intel Xeon processors) and a highly distributed configuration (8 nodes with 5-core Intel Xeon processors).

The remainder of this paper is organized as follows. Section~\ref{sec:motiv} presents the profiling benchmarks that motivate our architecture. 
Section~\ref{sec:anlys} details the StarDist backend analysis framework, while Section~\ref{sec:blkred} describes the underlying bulk-reduction communication substrate. Finally, Section~\ref{sec:eval} evaluates our system's performance.

\section{Motivation}
\label{sec:motiv}
\begin{figure}[t]
\centering
\begin{subfigure}[t]{0.21\textwidth}
\begin{lstlisting}
prop x;
prop y;
Edge e;
x.attachToGraph(g);
y.attachToGraph(g);
for v in g.nodes()
  for nbr in g.nbrs(v)
    e = get_edge(v,nbr);
    x.v = x.v + y.v;
\end{lstlisting}
\caption{Graph traversal}
\label{lst:gnbr}
\end{subfigure}
\hfill
\begin{subfigure}[t]{0.2\textwidth}
\begin{lstlisting}
prop x;
prop y;
x.attachToGraph(g);
y.attachToGraph(g);
for v in g.nodes()
  for nbr in g.nbrs(v)
    e = get_edge(v,nbr);
    e.x=<Min>(e.y+v.x);
\end{lstlisting}
\caption{Graph reduction op.}
\label{lst:gred}
\end{subfigure}
    \begin{subfigure}[c]{0.20\textwidth}
        \begin{lstlisting}
prop x;
prop y;
Edge e;
x.attachToGraph(g);
y.attachToGraph(g);
for v in g.nodes()
  for nbr in g.nbrs(v)
    (*@\textcolor{red}{e=forall(ed[v]==nbr);}@*)
    init_passiv_com(x);
    (*@\textcolor{red}{a = lockget(x[v]);}@*)
    term_passiv_com(x);
    init_passive_com(x);
    (*@\textcolor{red}{b = lockget(x[v]);}@*)
    term_passive_com(x);
    init_passiv_com(y);
    (*@\textcolor{red}{lockset(y[v], a+b);}@*)
    term_passiv_com(y)
        \end{lstlisting}
    \caption{Traversal target code}
    \label{lst:expnbr}
    \end{subfigure}
    \hfill
    \begin{subfigure}[c]{0.20\textwidth}
        \begin{lstlisting}
prop x;
prop y;
Edge e;
x.attachToGraph(g);
y.attachToGraph(g);
for v in g.nodes()
  for nbr in g.nbrs(v)
    (*@\textcolor{red}{e=forall(ed[v]==nbr);}@*)
    init_passiv_com(x);
    (*@\textcolor{red}{a = lockget(x[v]);}@*)
    term_passiv_com(x);
    init_passive_com(x);
    (*@\textcolor{red}{b = lockget(x[v]);}@*)
    term_passive_com(x);
    c = min(a,b);
    queue_msg(y[v], c);
(*@\textcolor{red}{all\_to\_all(queue);}@*)
init_passiv_com(y);
(*@\textcolor{red}{lockset(y[v], a+b);}@*)
term_passiv_com(y)
        \end{lstlisting}
    \caption{Reduction target code}
    \label{lst:expred}
    \end{subfigure}
\caption{StarPlat's current backend code generation on two patterns for graph traversal and updates}
\label{lst:generic_loop}
\end{figure}
\begin{figure}[t]
    \begin{subfigure}{0.20\textwidth}
        \begin{tikzpicture}
        \begin{axis}[
    ybar stacked,
    bar width=10pt,
    symbolic x coords={LV,OK,PK,RM,UM,WK},
    xtick=data,
    height = 3.6cm,
    x tick label style={
        rotate=45,
        anchor=east
    },
legend style={
    at={(0.5,-0.30)},   
    anchor=north,       
    legend columns=-1,  
        },
        ]
    \addplot [fill=oiBlue] table[x=G, y=NBR] {Figures/microbench1.txt};
    \addplot [fill=oiBlue!20] table[x=G, y=NBRC] {Figures/microbench1.txt};
    \legend{nbr, tot}
    \end{axis}
    \end{tikzpicture}
    \caption{\texttt{get-edge}}
    \label{fig:nbr}
\end{subfigure}
\hfill
    \begin{subfigure}{0.20\textwidth}
        \begin{tikzpicture}
        \begin{axis}[
    ybar stacked,
    bar width=10pt,
    symbolic x coords={LV,OK,PK,RM,UM,WK},
    x tick label style={
        rotate=45,
        anchor=east
    },
    xtick=data,
    height = 3.6cm,
    legend style={
    at={(0.5,-0.30)},   
    anchor=north,       
    legend columns=-1,  
        },
    ]
    \addplot [fill=oiBlue] table[x=G, y=RED] {Figures/microbench1.txt};
    \addplot [fill=oiBlue!20] table[x=G, y=REDC] {Figures/microbench1.txt};
    \legend{red, tot}
    \end{axis}
    \end{tikzpicture}
    \caption{\texttt{sync-reduce}}
    \label{fig:sred}
\end{subfigure}
\hfill
    \begin{subfigure}{0.20\textwidth}
        \begin{tikzpicture}
        \begin{axis}[
    ybar stacked,
    bar width=10pt,
    symbolic x coords={LV,OK,PK,RM,UM,WK},
    xtick=data,
    x tick label style={
        rotate=45,
        anchor=east
    },
    height = 3.6cm,
    legend style={
    at={(0.5,-0.30)},   
    anchor=north,       
    legend columns=-1,  
        },
        ]
    \addplot [fill=oiBlue] table[x=G, y=GET] {Figures/microbench1.txt};
    \addplot [fill=oiBlue!20] table[x=G, y=GETC] {Figures/microbench1.txt};
    \legend{get, tot}
    \end{axis}
    \end{tikzpicture}
    \caption{\texttt{get}}
    \label{fig:gcall}
\end{subfigure}
\hfill
    \begin{subfigure}{0.20\textwidth}
        \begin{tikzpicture}
        \begin{axis}[
    ybar stacked,
    bar width=10pt,
    symbolic x coords={LV,OK,PK,RM,UM,WK},
    xtick=data,
    x tick label style={
        rotate=45,
        anchor=east
    },
    height = 3.5cm,
    legend style={
    at={(0.5,-0.30)},   
    anchor=north,       
    legend columns=-1,  
        },
        ]
    \addplot [fill=oiBlue] table[x=G, y=ALL] {Figures/microbench1.txt};
    \addplot [fill=oiBlue!20] table[x=G, y=ALLC] {Figures/microbench1.txt};
    \legend{all, tot}
    \end{axis}
    \end{tikzpicture}
    \caption{Total time}
    \label{fig:all}
\end{subfigure}
\caption{Benchmarking of major phases of StarPlat}
\end{figure}
This work is built upon StarPlat~\cite{ref6}, a DSL for graph algorithms generating code for different backends: multi-core CPUs, many-core GPUs, and a distributed setup. It provides the \texttt{forall}, \texttt{iterBFS} and \texttt{fixedPointUntil} constructs for parallel operations across all supported backends.
The distributed backend for StarPlat generates code in the message passing model provided by OpenMPI~\cite{ref7}. For distributed code generation, whereas there is an optimization for global distributed variable updates, it is restricted to local variables declared outside a parallel construct. Graph property updates are not well optimized. 
StarPlat's MPI backend provides two alternatives for propagating updates to the graph property buffers distributed across partitions. When updating in bulk, programmers prefer using the \textit{reduction} construct, similar to the map-reduce~\cite{mapred} directive in GO; otherwise, direct assignment of values to properties leads to an RMA \texttt{PUT} call on the appropriate destination index of the RMA array which is appropriately locked and unlocked. The \textit{forall} syntax in StarPlat serves as an indication to the compiler for generating parallel code. While StarPlat effectively makes the generation of code in OpenMPI intuitive for the general programmer, we find the generated code lacking in performance. 
Two common patterns of graph loops and their generated codes in StarPlat are presented in Figure~\ref{lst:generic_loop}.
Figure \ref{lst:gnbr} shows a graph traversal pattern spanning all vertices and their neighbors performing a direct update of a graph property. The code generated in StarPlat in Figure \ref{lst:expnbr} involves a scan of the neighborhood of vertex v in search of the edge connecting v and nbr. Figure \ref{lst:gred} describes a reduction pattern which is compiled into Figure \ref{lst:expred}. This code initiates an access cycle in RMA and takes a lock on the RMA window for every element read. After reading, it unlocks the window and terminates the cycle. This degrades performance and makes the reduction message queue for updating in bulk, ornamental.

            
We profiled code generated by StarPlat for Push Relabel, subgraph isomorphism, and Single Source Shortest Paths using GNU's \texttt{gprof}~\cite{ref11} and RICE's \texttt{hpctoolkit}~\cite{ref12}. We found repeated patterns in graph traversal and graph property update that were either increasing cache miss rates or were injecting busy waits on distributed locks on RMA structures backing graph properties.
We also noticed that the edge list scan in the generated code, shown in Figure~\ref{lst:expnbr}, is used to retrieve an edge between two given vertices. (Boost Graph Library~\cite{ref31} design for storing edges as separate structures to adjacency lists.) Figure~\ref{fig:all} provides a breakdown of the execution time spent in the main phases of a distributed graph algorithm utilizing reduction in the generated SSSP algorithm. A reduction operation would comprise of \texttt{get} calls for operands, a \texttt{sync} call for propagating updates and the traversal across the buffer that is to be reduced.
 Figure~\ref{fig:gcall} indicates a large amount of time spent in \texttt{get} calls across a variety of naturally occurring graphs. The message queue provided by the reduction in Figure~\ref{lst:gred} accesses the property by initiating an RMA cycle and locking the entire window for every element that contributes to the message queue. The large stack depth these \texttt{get} calls require and the global locks on the windows cause a slowdown \textit{even when the property is present locally within the process address space}. We agree that the cost of \texttt{get} calls to foreign nodes cannot be avoided. However, if the block statement satisfies a certain \textit{reduction-exclusive property}, the subsequent \texttt{get} calls may be cached into the local address space.  The reduction-exclusive property is satisfied when a block statement has one reduction statement and the updates to participants in the reduction operation are unaffected by the ordering of updates.
Figure~\ref{fig:sred} indicates that a considerable amount of time is being spent synchronizing updates. StarPlat currently synchronizes updates at the end of the outermost forall block calling the reduction as is shown in Figure~\ref{lst:expred}. 
Figure~\ref{fig:all} indicates that SSSP (and more generally, other algorithms written in StarPlat) spend a super majority of their time in these phases.

We note in Figure \ref{iter:redroad} and in Figure~\ref{iter:redsoc} there is a cold start in SSSP-like algorithms where the amount of communication gradually increases to a peak and then fades into a long tail. Figure~\ref{iter:redroad} shows road networks in cold start even after 40 iterations. For social media networks where vertices have a higher average degree, the cold start is less prominent, however, the long tail exists. To avoid this, we introduce two strategies, a \texttt{short circuit} update locally for monotonic reduction operations, and a worklist driven analysis-transformation to fuse reduction-exclusive statements into \textit{pulses}, where we iterate between converging on our local subgraph and entering a global synchronization phase for interchanging outbound updates. 

\section{The Backend Analyzer}
\label{sec:anlys}
Figures~\ref{fig:nbr} and \ref{fig:sred} in Section~\ref{sec:motiv} indicate that the bottlenecks in StarPlat's distributed backend are in traversal patterns, in update-queuing and update-propagation. Our StarDist provides static analysis and transformations that optimize out redundant distributed global locks, provides transformations for more efficient graph traversal, and reduces the amount of communication required for convergence by fusing reduction operations into pulses. The analysis is coupled with a runtime for reduction operations (Section~\ref{sec:blkred}). We provide an excerpt from StarDist's Abstract Syntax Tree (AST) in Figure~\ref{fig:StarDist_bnf}. Reduction within the body of a \texttt{forall} statement, provide opportunities for our static analyzer to transform the program to propagate updates more efficiently. The \texttt{forall} statement's positioning with respect to divergence provides opportunities for both updating the elements and traversing the graph. In this section, we describe our optimization strategies for generating efficient distributed code.

\begin{figure}
\centering
   \begin{subfigure}[t]{0.20\textwidth}
    \begin{tikzpicture}
        \begin{axis}[xtick=data, ticks=none, xlabel = iterations, ylabel = number of updates,
        legend style={
        at={(0.75,1)},   
        anchor=north,       
        legend columns=2   
        },
             height = 4.5cm]            
            \addplot[oiBlue, mark=*, mark repeat = 3] table [x = 0, y=wk] {Experiments/iterations_scaled/wk};
            \addplot[oiOrange, mark=*, mark repeat = 3] table [x = 0, y=lv] {Experiments/iterations_scaled/lv};
            \addplot[oiVermillion, mark=*, mark repeat = 3] table [x = 0, y=ok] {Experiments/iterations_scaled/ok};
            \addplot[oiBluishGreen, mark=*, mark repeat = 3] table [x = 0, y=pk] {Experiments/iterations_scaled/pk.txt};
            \legend{WK, LV, OK, PK}
        \end{axis}
    \end{tikzpicture}
    \caption{Social media networks}
    \label{iter:redsoc}
\end{subfigure}
\hfill
\begin{subfigure}[t]{0.20\textwidth}
    \begin{tikzpicture}
        \begin{axis}[xtick=data, ticks=none, xlabel = iterations, ylabel = number of updates,
        legend style={
        at={(0.357,1)},   
        anchor=north,       
        legend columns=2   
        },
             height = 4.5cm]            
            \addplot[oiSkyBlue, mark=*, mark repeat = 3] table [x = 0, y=gm] {Experiments/iterations_scaled/gm.txt};
            \addplot[oiReddishPurple, mark=*, mark repeat = 3] table [x = 0, y=us] {Experiments/iterations_scaled/us};
            \legend{GM, US}
        \end{axis}
    \end{tikzpicture}
    \caption{Road networks}
    \label{iter:redroad}
\end{subfigure}
    \caption{\#Updates across SSSP iterations in StarPlat}
    \label{fig:placeholder}
\end{figure}

\begin{figure}[t]
\centering
\begin{bnf*}
\bnfprod{forall}
    {\bnfpn{forall signature} \bnfpn{block}}\\
\bnfprod{block}
    {\bnfpn{stmt} \bnfpn{stmt} \bnfor \bnfpn{stmt}}\\
\bnfprod{stmt}{\bnfpn{reduction}}\\
\bnfprod{reduction}{\bnfpn{<}\bnfpn{op}\bnfpn{>}\bnfpn{Expr} }\\
\bnfprod{Expr}{\bnfpn{Expr}\bnfpn{Expr}}\\
\bnfprod{Expr}{\bnfpn{propE} \bnfor \bnfpn{arthE} \bnfor \bnfpn{procE} \bnfor \bnfpn{ID}}\\
\end{bnf*}
\caption{StarDist grammar rules for optimizations}
\label{fig:StarDist_bnf}
\end{figure}

\subsection{Analysis and Optimizations}
\begin{figure*}[t]
\centering
\begin{subfigure}[c]{0.20\textwidth}
\begin{lstlisting}
prop dist;
prop weight;
Edge e;
dist.attachToGraph(INTMAX);
dist.lockset(src,0);
g.reduction.push(src);
while g.reduction()
  for v in g.reduction()
    for nbr in g.nbrs(v)
      e=get_edge(v,nbr);
      t=v.dist+e.weight
      nbr.dist = <Min>(v.dist + t)
\end{lstlisting}
\caption{Single Source Shortest Paths in StarDist.}
\label{lst:ssspsource}
\end{subfigure}
\hfill
\begin{subfigure}[c]{0.20\textwidth}
\begin{lstlisting}
prop dist;
prop weight;
Edge e;
dist.attachToGraph(INTMAX);
dist.lockset(src,0);
g.reduction.push(src);
while g.reduction()
 for v in g.reduction()
  for nbr in g.nbrs(v)
    e = next_edge(v);
    nbr=other_end(v,e);
    t = v.dist + e.weight
    queue(nbr, t);
  sync();
\end{lstlisting}
\caption{After neighbourhood traversal analysis.}
\label{lst:ssspnbrpass}
\end{subfigure}
\hfill
\begin{subfigure}[c]{0.20\textwidth}
\begin{lstlisting}
prop dist;
prop weight;
Edge e;
d.attachToGraph(INTMAX);
d.lockset(src,0);
g.reduction.push(src);
while g.reduction()
  for v in g.reduction()
    for nbr in g.nbrs(v)
      e = next_edge(v);
      nbr=otherend(v,e);
      if v is local
        x = d.arr[v];
        if e is local
          y = w.arr[v];
        else
        y=w.lockget(e);
      else
        x=d.lockget(v);
    queue(nbr, x + y);
  sync();
\end{lstlisting}
\caption{After get call optimizations.}
\label{lst:ssspredpass}
\end{subfigure}
\hfill
\begin{subfigure}[c]{0.20\textwidth}
\begin{lstlisting}
prop dist;
prop weight;
Edge e;
d.attachToGraph(INTMAX);
d.lockset(src,0);
g.red.push(src);
while g.red()
 while g.red()
  for v in g.red()
   for nbr in g.nbrs(v)
    e=next_edge(v);
    nbr=otherend(v,e);
    if v is local
     x=d.arr[v];
      if e is local
        y=w.arr[v];
      else
        y=w.lockget(e);
    else
      x=d.lockget(v);
    queue(nbr, x + y);
  swap();
 sync();
\end{lstlisting}
\caption{After pulse optimization.}
\label{lst:pulsepass}
\end{subfigure}
\caption{Analysis passes on SSSP written in StarDist}
\end{figure*}
 Figure~\ref{lst:ssspsource} shows the source code for SSSP in StarDist. Upon undergoing the traversal analysis, it is updated as in Figure~\ref{lst:ssspnbrpass} where the vertex traversal is replaced by an edge traversal. \texttt{get} call analysis leads to the code in Figure~\ref{lst:ssspredpass} and finally, since this code segment utilizes a monotonic reduction operation, we get the code in Figure \ref{lst:pulsepass}, egmented into pulses that aggregate maximum communication.
 
 We note that a \texttt{forall} construct performing a reduction on graph properties does not result in the property being updated until after the construct completes. If the property is read in the body of the \texttt{forall}, the previous value is expected to be obtained. Updates to the property buffers that do not exhibit RAW, WAR, or WAW data dependencies within the block statement, if resident locally, may be updated. Updates to the property buffers exhibiting data dependencies within a monotonic reduction operation may also be updated locally, as the reduction operator would converge to the same value regardless. Propagating updates to the local subgraph significantly reduces the message queue size, thus impacting the overall time spent in communication.
 A monotonic operation is a non-decreasing or a non-decreasing function. This property of the monotonic operator allows for convergence to the same value regardless of repeated operations given that all the operands perform the operation at least once. A monotonic reduction is a reduction operation where the operator obeys monotonicity and repeated operations do not result in a different result.


\subsection{Graph Traversal Static Analysis}
To attach and maintain edge properties, StarDist records edges as structures separate from the vertices. This design is similar to that of \texttt{boost}'s graph library~\cite{ref31}. 
In a pattern as in Figure~\ref{lst:gnbr}, the generated code does not recognize that it is a complete neighborhood traversal of the subgraph vertices owned locally. This results in a search for the edge between \texttt{(v,nbr)} as shown in Figure~\ref{lst:expnbr} and adds the \texttt{O(degree(v))} traversal for every \texttt{(v,nbr)} combination.
We note that the MPI runtime does not store the neighborhood of a vertex in its Compressed Sparse Representation (CSR) in a predetermined order, but rather in the sequence in which the edges are encountered in the edge list. 
The \texttt{forall} vertex traversal loop directive in \texttt{StarDist} symbolizes parallel traversal and therefore data regions modified by the operation would be invariant to the order in which updates happen. When a vertex traversal is executed in the outermost \texttt{forall} statement that is not within a divergent block statement (we call a block statement executed by some of the processes in the distributed system and not all, a divergent block statement), we traverse only the vertices that exist locally. Since MPI follows the SPMD (Single Program Multiple Data) model, when this vertex traversal is in a non-divergent block, the participating processes collectively traverse all the graph vertices. This gives us semantic confirmation that the loop variable vertices (vertex \texttt{v} in Figure~\ref{lst:generic_loop}) are local and their neighborhood information (available in the CSR) must also be available locally. The symbol table on StarPlat stores information about the vertex traversal and neighborhood traversal. StarDist performs an auto-transformation that if it observes a neighborhood traversal within a vertex traversal, and there is an access to an edge-structure comprising of vertices belonging to the iterator variables in the \texttt{forall} signatures, it shuffles the vertex traversal with an edge traversal and replaces the \texttt{nbr} vertex with the other vertex of the edge. This eliminates the need to scan for vertices in the edge list when getting the edge structure and replaces it with an \texttt{O(1)} random access into the edge to find the other vertex.

\subsection{Update Queue and Propagation Analysis}
StarDist, like StarPlat, does not maintain ghost nodes (Riley et al.~\cite{ref32}), but updates properties in foreign address space either by queuing them into a \textit{reduction message queue} for later synchronization or propagating them immediately through an RMA \texttt{put} call.
We introduce a reduction-exclusive statement which forms the basis of our update propagation analysis-transformation passes.
The analysis explained in this subsection and a bulk reduction component in Section~\ref{sec:blkred} provide enhancements that take advantage of the reduction-exclusive pattern for commutative and monotonic reduce operations. 
\begin{definition}
    Two statements $S_1$ and $S_2$, operating on the same property buffer sets $E_1$ and $E_2$ are \textbf{reduction dependent} if $E_1 \cap E_2 \ne \phi$ and atleast one of $S_1$ and $S_2$ are a reduction statement with a non-monotonic reduction operation. Therefore, if $S_1$ is performing a reduction operation on a property buffer and $S_2$ modifies elements in that buffer, we say that $S_1$ and $S_2$ are reduction-dependent. We define a \texttt{block} statement to be \textbf{reduction exclusive} if there is exactly one reduction statement $S_i$ within the block statement and no other statements in the block statement are reduction-dependent with $S_i$.
    \label{def:reduction-exclusive}
\end{definition}

\begin{definition}
If $S_1$ and $S_2$ access element sets $E_1$ and $E_2$, while $S_1$ and $S_2$ belong to a local vertex traversal in a non-divergent outermost \texttt{forall}, and $E_1 \cap E_2 = \{v\}$ where $v$ is the iterator for the local node traversal, we call $S_1$ and $S_2$ to be \textbf{global update invariant}.
\label{def:glob-inv}
\end{definition}


\begin{definition}
We define exhausting all intra-process communication before entering a period of intra-process computation as a \textbf{pulse}.
\label{pulse}
\end{definition}

\begin{lemma}
A block statement within a reduction-exclusive statement is also a reduction-exclusive statement.
\label{extension lemma}
\end{lemma}

\begin{proof}
Given a  reduction exclusive statement which contains statements $X$ as children in the AST, then any two statements $X_i$ and $X_j$ belonging to $X$ are not reduction independent. So, from definition \ref{def:reduction-exclusive}, $X$ is reduction exclusive.
\end{proof}

Pulse (Definition~\ref{pulse}) is an extended, more abstract form of that in Goldberg and Tarjan's push-relabel algorithm~\cite{ref8}, where delayed and aggregated communication was proven beneficial for distributed push-relabel.

\paragraph{get call bypass} Queuing of updates in \texttt{reduction message queues} often follows \texttt{get} calls for values of certain node or edge properties of the graph that are backed by the OpenMPI RMA~\cite{ref7} framework as shown in Figure~\ref{lst:gred}. The \texttt{get} calls for RMA are expensive even for local vertices, due to RMA cycle initiation overhead, global locking of the RMA window, and the stack depth involved in the \texttt{get} call.
OpenMPI does not internally optimize subsequent RMA calls by keeping the RMA phases active~\cite{ref7}. We observed that keeping the RMA window with a shared lock also did not optimize the \texttt{get} calls.
Our analysis expands the expression within the reduction operator and creates a nested conditional block. For every block, the residence of the vertex or the edge is queried and two different queuing statements are generated as shown in Figure~\ref{lst:expred}.
We use Lemma~\ref{extension lemma} and our analysis for determining reduction-exclusive statements (Definition \ref{def:reduction-exclusive}) to argue that \texttt{get} calls within reduction-exclusive statements can safely bypass RMA directives and dereference the local memory of an RMA window, as updates to the data undergoing reduction are not going to be propagated until the end of the ongoing pulse, assuring that WAW, RAW, and WAR dependencies are not possible.

\paragraph{global update invariant}
Figure~\ref{lst:expnbr} reads and updates properties of only vertex $v$, where $v$ is within the non-divergent vertex traversal and, therefore, is locally resident. $v$ is also not read by anyone. and therefore, we guarantee that it is a global invariant (Definition~\ref{def:glob-inv}). Global-invariants do not incur RAW, WAR, or WAW dependencies. We can prove by contradiction that if there was a dependence, then $E_1 \cap E_2 \supset {v}$ and $v$ fails the Definition~\ref{def:glob-inv}.
After the safety analysis pass, we simply pass a fast path update directive to the StarDist runtime, which results in an immediate update.

\paragraph{Pulse formation}
Using Lemma~\ref{extension lemma}, we extend the reduction statement in StarPlat to form pulses out of reduction-exclusive statements when the reduction operation is monotonic, effectively \textit{fusing} multiple temporally separated reduction operations. While this introduces WAW, RAW, and WAR dependencies, we inject a convergence factor by introducing a worklist. Any vertex that got assigned a value closer to the bound in the partial ordering is pushed into the worklist for processing in the next round.
Lemma~\ref{extension lemma} allows us to envelope the reduction statement into a fixedpoint iteration until the worklist is empty. 
A pulse can be constructed only when the updates are commutative, associative, and monotonic. Within the pulse, local updates are propagated, and foreign updates are queued. Once no more local updates are possible, a global update propagation phase occurs. This is repeated until convergence (empty worklist).

\section{Bulk Reduction Substrate}
\label{sec:blkred}
\begin{figure}[t]
  \centering
  \begin{subfigure}[t]{0.22\textwidth}
    \graphdd
    \caption{StarPlat's MPI headers}
    \label{fig:spheader}
  \end{subfigure}
  \hfill
  \begin{subfigure}[t]{0.24\textwidth}
      \graphndd
      \caption{StarDist's MPI headers}
      \label{fig:sdheader}
  \end{subfigure}
  \caption{MPI bulk reduction runtime}
  \label{fig:bulk_red}
\end{figure}
 StarPlat utilizes its MPI runtime library provided alongside for communication procedures. The runtime maintains the distributed graph structure and has implementations for the reduction operation, vertex (np) and edge properties(ep), and the RMA data structures supporting all the operations 
(Figure~\ref{fig:spheader}). 
 In StarDist, we introduce a bulk reduction(blkRed) substrate to the runtime for RMA backed reduction, message volume reduction, and fast-path update optimizations. We couple our runtime with a singleton global structure that spans across the MPI runtime and collects data through multiple sub-components as shown in Figure~\ref{fig:sdheader}. 
 The bulk reduction runtime provides different levels of optimizations based on the reduction operation passed on by the AST. For commutative and monotonic reductions, StarDist is able to reduce the reduction message queue for outbound updates and is able to set values of properties locally resident through a fast path. 
 Pulses (Definition~\ref{pulse}) are only formed if the StarDist compiler's data dependency constraints are satisfied and the reduction operation is monotonic and commutative. These reduction patterns reduce communication, but stress the reduction queue based on the degree of fusion provided by the backend analyzer. We use passive MPI RMA synchronization directives for batch propagating updates, and provide configurations to adjust the batch size to make the active segments of the substrate structures resident within the lowest level cache.

\subsection{Reduction Message Queue}
In our distributed execution model, much like StarPlat~\cite{ref5}, graph properties (such as vertex distances or PageRank values) are maintained as partitioned distributed arrays maintained using one sided communication directives provided in MPI. Each process holds exclusive ownership over a specific sub-segment of the global state depending on the subgraph the process owns. We implement a reduction message queue to manage the irregular, fine-grained updates to these distributed arrays generated during graph traversal. 
\paragraph{design}
Structurally, this queue is designed as a linked list of contiguous memory allotments. Although buffering irregular graph updates into contiguous memory is a well-established technique to amortize communication overhead~\cite{ref4,powergraph}, relying on the standard dynamic resizing of arrays for this buffer can introduce unpredictable latency spikes in the critical path due to \texttt{O(NlogN)} memory reallocation and copying. 
We utilize a linked list of contiguous memory, so that memory allocation overhead is strictly \texttt{O(1)}. This ensures that while updates are being queued, execution time is not hampered when copying buffers as would happen for amortized dynamic arrays.
As outlined in Algorithm \ref{algo:addred}, the reduction message queue acts as the central routing layer for all generated updates, regardless of their final destination. When an update x targets a vertex (or edge) u resident on process i (which may be the local process or a remote one), the substrate multiplexes the (u,x) tuple directly into the active memory chunk assigned to the i-th partition. While the generated code performs irregular accesses, this queueing ensures that all updates are spatially grouped together by their target distributed array segment well before any network activity occurs.
During the queueing phase of the reduction, the local process strictly writes generated updates into the message queue fragments. Once the computation phase concludes, all local execution halts, and the synchronization phase begins. For remote partitions, the network substrate traverses the linked list, by doing an RMA \texttt{put} of one batch in the queue per destination rank at a time. This coagulates small, frequent updates into sizeable updates for the underlying RMA calls to be advantageous. 
\begin{algorithm}
    \caption{ADD TO REDUCTION WITH QUEUE SHORTENING}
    \label{algo:addred}
    \KwIn{vertex $v$, val $val$, REDOP $red\_op$}
        \If {$red\_op$ is commutative}{
            \If {$red\_op$ is monotonic and $v$ is local}{
                prop[$local\_index(v)$] = $val$;
                return;
            }
            \If{idx in idx\_cache[$owner(v)$]}{
                entry = idx\_cache[idx]\;
                *entry = $red\_op$(val)\;
                return;
            }
        }
        \If {buffer[$owner(v)$] is full}
        {
            buffer[$owner(v)$]->next = malloc(buffer\_sz/$world\_size$)\;
            buffer[$owner(v)$] = buffer[$owner(v)$]->next\;
            index[$owner(v)$] = 0\;
        }
        buffer[$owner(v)$][$INDEX(owner(v))$] = v\;
        buffer[$owner(v)$][$INDEX(owner(v)+1$] = val\;
        index[$owner(v)$] = index[$owner(v)$] + 2\;
\end{algorithm}
\paragraph{queue shortening}
Updates propagated via a reduction call are not needed within a reduction-exclusive statement.
If the reduction operation is monotonic, the substrate will \textit{short-circuit} the update to $v$ if $v$ is resident locally. A monotonic operation with a fixed point convergence, that repeats operations redeem the same result. Therefore reduction-exclusive statements within pulses or when attached to a worklist \cite{mfa}, merging together partially do not compromise correctness.

\subsection{Synchronizing Reduction}
We synchronize reduction in multiple passes, in a separate global synchronization phase during which all participating processes enter the synchronization segment. The reduction RMA window is divided into equal batch sized parts that act as recipient windows for one batch from every process. Algorithm~\ref{algo:addred} ensures the reception window is filled only with updates for properties resident within receiver's subgraph. At the completion of one batch of synchronization, the local updates are propagated. The synchronization and update procedure is detailed in algorithm~\ref{algo:bulk_sync}.
The synchronize-update phase happens in lockstep. Therefore, a passive shared lock is taken with the set $no\_check\_needed$ flag on all windows. This way of locking has been shown to be 49\% more efficient in \cite{RMA}. Every process traverses the list of arrays in the reduction queue, one destination at a time. 
Once the movement is completed, the processes read their regions of the RMA window and update the respective node property.
The transfer happens in chunks until all the updates are propagated. StarPlat's synchronize reduction uses an all-to-all model that congests the network, and in turn, proves difficult to scale.
\begin{algorithm}
    \caption{BULK SYNC algorithm}
    \label{algo:bulk_sync}
    bufptr = buffer->head\;
    \While{buffer}
    { 
    mpi\_win\_lock\_all(shared)\;
    
    \For {rank in world} 
    {
            $rma\_wind \gets buftpr[rank]$\;
    }
    mpi\_win\_unlock\_all(shared)\;
    barrier\; \
    \For {index in buffersize}
    {
        idx = rma\_wind[index]\;
        val = rma\_wind[index+1]\;
        redop(idx, val)\;
    }
    bufptr = buffer->next\;
    }
\end{algorithm}

\section{Experimental Evaluation}
\label{sec:eval}
\subsection{Experimental Setup} 
Our experimental comprises two distinct setups. The first comprises three nodes with each node comprising an Intel Xeon Gold 6142 processor with 32 cores and 192 GB of RAM. Each node has 2 NUMA regions comprising 16 cores. The three nodes are interconnected via a 10 Gbps RDMA infiniband. The 32 processes per node communicating with each other congests the operating system network stack.
Our second setup is composed of 8 nodes connected through a 10 Gbps RDMA infiniband. We occupy 5 cores per node and use this setup to demonstrate StarDist generated distributed code's scaling abilities. 
Our second setup results in a more complex network topology that represents a noisy environment. We perform a comparative study, strong scaling experiments, and weak scaling experiments as part of our evaluation. We utilize the \texttt{scatter} PBS directive to ensure maximum cross-node placement.
We perform strong scaling on our first setup with 3 nodes and 32 processors. Our weak scaling study is performed on our second setup whereas our comparative study is done on both the setups.

\paragraph{Baselines}
\texttt{DRONE}~\cite{DRONE} and \texttt{d-Galois}~\cite{Galois,Galois2} are subgraph-centric distributed graph algorithm frameworks which distribute graphs into partitions containing master and mirror vertices. They synchronize the master and mirror nodes in phases, usually after propagating all updates locally (much like StarDist \textit{pulses}). \texttt{d-Galois} is a well established framework and utilizes the Gluon async~\cite{ref1} runtime for synchronizing master and mirror nodes. 
While \texttt{d-Galois} is a purely peer-to-peer framework, DRONE follows a master-worker design with the master process providing bookkeeping and synchronization initiatives for the worker processes.
While writing algorithms in StarDist is easy, we have found doing so in other frameworks for distributed graph algorithms require manual coding of the communication pattern with framework compliant function pointers written for them. SSSP, WCC, and PR are  the most common implementations across these frameworks and are therefore chosen for our comparative study. 
We conducted five iterations per experiment and reported the average.



\paragraph{DATASET}
We perform our experiments in the following test suite that contains a mixture of social media and road network graphs. For SSSP, we have added random weights of $[1,100]$ to all the unweighted graphs. We utilize RMAT~\cite{ref30} generated graphs for weak scaling analysis of our algorithms and real world graphs (Table~\ref{tab:graphs}) in the suite for comparison with the prior art and scaling analysis. Our real world graphs have been observed to exhibit the power-law property where there are only a few vertices with high degree.
Our RMAT graphs are generated using Stanford's SNAP library~\cite{snap} with parameters A, B, and C having values 0.5, 0.3, and 0.09, leaning the graphs more toward road networks, which are time-consuming to solve in parallel programming models due to the slow growth of the active vertices worklist.

\begin{table}[h]
\centering
\begin{tabular}{|l|c|r|r|}
\hline
\textbf{Graph Name} & \textbf{Abbrev.}  & \textbf{\#Nodes} & \textbf{\#Edges} \\
\hline
soc-twitter-2010    & TW    & 21,297,772 & 265,025,809 \\
soc-sinaweibo       & SW    & 58,655,849 & 261,321,071 \\
com-Orkut           & OK    & 3,072,441  & 234,370,166 \\
Wikipedia-links(ru) & WK    & 3,370,462  & 93,373,056  \\
soc-LiveJournal1    & LJ    & 4,847,571  & 68,993,773  \\
soc-pokec           & PK    & 1,632,803  & 30,622,564  \\
USARoadNet          & US    & 23,947,347 & 57,708,624  \\
germany-osm         & GM    & 11,548,845 & 24,738,362  \\
RMAT                & RM    & 16,777,216 & 87,654,320  \\
Uniform Random      & UR    & 10,000,000 & 80,000,000  \\
\hline
\end{tabular}
\caption{Graph datasets used in evaluation}
\label{tab:graphs}
\end{table}

\subsection{Scaling Experiments}
\begin{figure}[t]
    \begin{subfigure}{0.20\textwidth}
    \begin{tikzpicture}
        \begin{axis}[grid, xlabel=procs, ylabel=time, height=4.0cm,
        legend to name=sharedlegend,
        font=\footnotesize,
        legend style={legend columns=4} 
        ]
            \addplot[oiBlue, mark=*] table[x=np,y=time]{Experiments/experiment_sssp_scale/exp1_proced0pokecwt.txt};
            \addlegendentry{PK}
            \addplot[oiOrange, mark=*] table[x=np,y=time]{Experiments/experiment_sssp_scale/exp1_proced0livjournalwt.txt};
            \addlegendentry{LJ}
            \addplot[oiBluishGreen, mark=*] table[x=np,y=time]{Experiments/experiment_sssp_scale/exp1_proced0orkutudwt.txt};
            \addlegendentry{OK}
            \addplot[oiReddishPurple, mark=*] table[x=np,y=time]{Experiments/experiment_sssp_scale/exp1_proced0twitterwt.txt};
            \addlegendentry{TW}
            \addplot[oiBlue, dashed, mark=*] table[x=np,y=time]{Experiments/experiment_sssp_scale/exp1_proced1pokecwt.txt};
            \addlegendentry{pulse-PK}
            \addplot[oiOrange, dashed, mark=*] table[x=np,y=time]{Experiments/experiment_sssp_scale/exp1_proced1livjournalwt.txt};
            \addlegendentry{pulse-LJ}
            \addplot[oiBluishGreen, dashed, mark=*] table[x=np,y=time]{Experiments/experiment_sssp_scale/exp1_proced1orkutudwt.txt};
            \addlegendentry{pulse-OK}
            \addplot[oiBlack, mark=*] table[x=np,y=time]{Experiments/experiment_sssp_scale/exp1_proced0wikiwt.txt};
            \addlegendentry{WK}
            \addplot[oiBlack, dashed, mark=*] table[x=np,y=time]{Experiments/experiment_sssp_scale/exp1_proced1wikiwt.txt};
            \addlegendentry{pulse-WK}
            \addlegendimage{oiReddishPurple, dashed, mark=*}
            \addlegendentry{pulse-TW}
            \addlegendimage{oiVermillion, mark=*}
            \addlegendentry{GM}
            \addlegendimage{oiVermillion, dashed, mark=*}
            \addlegendentry{pulse-GM}
            \addlegendimage{oiSkyBlue, mark=*}
            \addlegendentry{US}
            \addlegendimage{oiSkyBlue, dashed, mark=*}
            \addlegendentry{pulse-US}
        \end{axis}
    \end{tikzpicture}
    \caption{SSSP on social graphs}
    \label{scale:ssspsoc}
    \end{subfigure}
    \hfill
    \begin{subfigure}{0.20\textwidth}
    \begin{tikzpicture}
        \begin{axis}[grid, height=4.0cm,xlabel=procs,ylabel=time,
                legend style={
        at={(0.25,1)},   
        anchor=north,       
        legend columns=2   
        }]

            \addplot[oiVermillion, mark=*] table[x=np,y=time]{Experiments/experiment_sssp_scale/exp1_proced0germanyudwt.txt};
            \addplot[oiVermillion, dashed, mark=*] table[x=np,y=time]{Experiments/experiment_sssp_scale/exp1_proced1germanyudwt.txt};
        \end{axis}
    \end{tikzpicture}
    \caption{SSSP on GM}
    \label{scale:sssprd}
    \end{subfigure}
    \hfill
\begin{subfigure}{0.20\textwidth}
\begin{tikzpicture}
        \begin{axis}[grid, height=4.0cm, xlabel=procs, ylabel=time, 
        legend style={
        at={(0.5,-0.14)},   
        anchor=north,       
        legend columns=2   
        }]
            \addplot[oiOrange, mark=*] table[x=nprocs,y=time]{Experiments/strong_scale_wcc/nonfused/liv.tex};
            \addplot[oiOrange, dashed, mark=*] table[x=nprocs,y=time]{Experiments/strong_scale_wcc/fused/liv.tex};
            \addplot[oiBlue, mark=*] table[x=nprocs,y=time]{Experiments/strong_scale_wcc/nonfused/pok.tex};
            \addplot[oiBlue, dashed, mark=*] table[x=nprocs,y=time]{Experiments/strong_scale_wcc/fused/pok.tex};
            \addplot[oiBluishGreen, mark=*] table[x=nprocs,y=time]{Experiments/strong_scale_wcc/nonfused/orkut.tex};
            \addplot[oiBluishGreen, dashed, mark=*] table[x=nprocs,y=time]{Experiments/strong_scale_wcc/fused/orkut.tex};
            \addplot[oiBlack, mark=*] table[x=nprocs,y=time]{Experiments/strong_scale_wcc/nonfused/wiki.tex};
            \addplot[oiBlack, dashed ,mark=*] table[x=nprocs,y=time]{Experiments/strong_scale_wcc/fused/wiki.tex};
        \end{axis}
    \end{tikzpicture}
    \caption{WCC on social graphs}
\end{subfigure}
\hfill
\begin{subfigure}{0.20\textwidth}
\begin{tikzpicture}
        \begin{axis}[grid,height=4.0cm,xlabel=procs,ylabel=time]
            \addplot[oiVermillion, mark=*] table[x=nprocs,y=time]{Experiments/strong_scale_wcc/nonfused/germ.tex};
            \addplot[oiVermillion, dashed, mark=*] table[x=nprocs,y=time]{Experiments/strong_scale_wcc/fused/germ.tex};
            \addplot[oiSkyBlue, mark=*] table[x=nprocs,y=time]{Experiments/strong_scale_wcc/nonfused/usa.tex};
            \addplot[oiSkyBlue, dashed, mark=*] table[x=nprocs,y=time]{Experiments/strong_scale_wcc/fused/usa.tex};
        \end{axis}
    \end{tikzpicture}
    \caption{WCC on road graphs}
\end{subfigure}
\hfill
\begin{subfigure}{0.20\textwidth}
\begin{tikzpicture}
        \begin{axis}[grid,height=4.0cm,xlabel=procs,ylabel=time]
            \addplot[oiOrange, mark=*] table[x=nprocs,y=time]{Experiments/strong_scale_pr/liv.txt};
            \addplot[oiBluishGreen, mark=*] table[x=nprocs,y=time]{Experiments/strong_scale_pr/orkut.txt};
            \addplot[oiBlue, mark=*] table[x=nprocs,y=time]{Experiments/strong_scale_pr/pok.txt};
            \addplot[oiSkyBlue, mark=*] table[x=nprocs,y=time]{Experiments/strong_scale_pr/rmat.txt};
            \addplot[oiBlack, mark=*] table[x=nprocs,y=time]{Experiments/strong_scale_pr/wiki.txt};
        \end{axis}
    \end{tikzpicture}
    \caption{PR on social networks}
\end{subfigure}
\hfill
\begin{subfigure}{0.20\textwidth}
\begin{tikzpicture}
        \begin{axis}[grid,height=4.0cm,xlabel=procs,ylabel=time]
            \addplot[oiVermillion, mark=*] table[x=nprocs,y=time]{Experiments/strong_scale_pr/germ.txt};
        \end{axis}
    \end{tikzpicture}
    \caption{PR on road networks}
\end{subfigure}
\ref{sharedlegend}
    \caption{Strong scaling of reduction with fusion vs. reduction without fusion on social media networks and road networks for SSSP, WCC, and PR}
    \label{comp:fusion}
\end{figure}
\paragraph{Strong Scale:}
For our strong scaling experiments, we run SSSP, WCC, and PR generated by StarDist on our compact setup and increase the number of processes from 10 to 90.
The execution time in our DSL test suite showed good scaling for SSSP, as shown in Figure~\ref{scale:ssspsoc} and in Figure~\ref{scale:sssprd}. While keeping initiating one MPI process per hardware core, we observe an exponential dip in execution time intra-node and a linear dip in execution time inter-node. While utilizing RMA calls for synchronizing reductions would be faster, when scaled across processes scattered throughout NUMA, we should have seen an increase. By increasing the number of processes, only the amount of intra-node work is reduced. This supports our earlier claim (Section~\ref{sec:motiv}) that synchronization overheads may not always be the main bottleneck in distributed graph algorithms. The queuing of updates into the reduction message queue itself is expensive due to poor locality and a large number of insert operations into the message queue. We compare SSSP generated by StarDist fused into pulses with reduction short-circuiting, and the code with only reduction short-circuits. For SSSP with pulse fusion analysis, we notice little to no improvement on social media networks. On road networks, without pulse fusion, performance on GM deteriorates with scale. Pulse fusion not only improves the performance on GM, but makes it scale as well. On the other hand, for US, pulse fusion degrades performance (Table~\ref{tab:comp}).
 Page Rank does not utilize only the reduction operation. PR code in StarPlat mimics distributed Galois's logic where deltas are calculated in the first pass followed by another phase scaling them down and adding them to the Page Rank array.
DRONE did not handle the divide-by-zero exception that is encountered if a vertex's outdegree is 0. We added that check, and nothing else. We ignored adding such delta to the vertices keeping in line with d-Galois' implementation. PR did not pass analysis for pulse formation due to the summation reduction operator. However, due to the algorithm reducing to a separate buffer, short-circuiting of updates happened safely.


\begin{figure*}
\begin{subfigure}{0.33\textwidth}
\begin{tikzpicture}
        \begin{axis}[height=4.0cm, ymin=0, grid, xlabel=nodes,ylabel=time,legend to name=sharedlegend2,
        font=\footnotesize,
        legend style={legend columns=2}]
            \addplot[oiBlue, mark=*] table[x=nodes,y=time]{Experiments/weak_scale/wcc/WCC.txt};
            \addlegendentry{DSEDG}
            \addplot[oiBlue, dashed, mark=*] table[x=nodes,y=time]{Experiments/weak_scale/wcc/WCC_pulse.txt};
            \addlegendentry{fused DSEDG}
        \end{axis}
    \end{tikzpicture}
    \caption{WCC on RMAT graphs}
    \label{scale:weakwcc}
\end{subfigure}
\begin{subfigure}{0.33\textwidth}
\begin{tikzpicture}
        \begin{axis}[height=4.0cm, ymin=0, grid,xlabel=nodes,ylabel=time]
            \addplot[oiBlue, mark=*] table[x=nodes,y=time]{Experiments/weak_scale/sssp/sssp.txt};
            \addplot[oiBlue, dashed, mark=*] table[x=nodes,y=time]{Experiments/weak_scale/sssp/sssp_pulse.txt};
        \end{axis}
    \end{tikzpicture}
    \caption{SSSP on RMAT graphs}
    \label{scale:weaksssp}
\end{subfigure}
\begin{subfigure}{0.33\textwidth}
\begin{tikzpicture}
        \begin{axis}[height=4.0cm, ymin=0, grid,xlabel=nodes,ylabel=time]
            \addplot[oiBlue, mark=*] table[x=procs,y=avg]{Experiments/weak_scale/pr/pr.txt};
        \end{axis}
    \end{tikzpicture}
    \caption{PR on RMAT graphs}
    \label{scale:weakpr}
\end{subfigure}
\ref{sharedlegend2}

\caption{Weak scaling across 8 inter-connected nodes}
\textit{Note: DSEDG = DisteEdge; fused DSEDG = DisteEdge with pulse optimization}

\end{figure*}

\paragraph{Weak Scaling}
For weak scaling, we keep the problem size per core constant at 8,000,000 edges per core while analyzing performance across multiple cores. We used our second setup with 8 nodes for this experiment as we wanted to capture delays introduced due to large communication costs. Figure~\ref{scale:weakwcc} demonstrates weak scaling of WCC in RMAT-generated graphs on 8 nodes. We notice the bottleneck introduced by bandwidth taking effect and the time taken increasing. Figure~\ref{scale:weakpr} demonstrates the same for PR. PR remained fairly consistent with increasing number of cores. Figure~\ref{scale:weaksssp} shows good weak scaling for 6 nodes and a spike at the 8th node. The RMA graphs for weak scaling SSSP were modified so as to make the entire graph connected. This was done to make sure all the graphs offered proportional work for SSSP. We found DRONE's partitioning algorithm crashing on our generated RMAT graphs and d-Galois' partitioning algorithm kept timing out on a 30 minute timer. So, we could not include them as part of our weak scaling analysis.

\subsection{Comparison with Prior Art}
\begin{figure}[t]
\begin{subfigure}[t]{0.20\textwidth}
    \begin{tikzpicture}
        \begin{axis}[symbolic x coords={OK, WK, LJ, PK, RM, UR},xtick=data, height=5.5cm,
        legend to name=sharedlegend3,
        legend style={
            legend columns=-1,
            font=\footnotesize,
        },
            x tick label style={
        rotate=45,
        anchor=east
    }, ybar=0pt, ymax=60, height=4.0cm, bar width=0.1cm, ylabel=time]
            \addlegendimage{ybar,fill=oiBluishGreen,draw=black, area legend}
            \addlegendentry{DSEDG}
            \addlegendimage{ybar,fill=oiSkyBlue,draw=black, area legend}
            \addlegendentry{DSEDGPLS}
            \addlegendimage{ybar,fill=oiVermillion,draw=black, area legend}
            \addlegendentry{DRONE}
            \addlegendimage{ybar,fill=oiOrange,draw=black, area legend}
            \addlegendentry{GAL}
            \addlegendimage{ybar,fill=oiBlue,draw=black, area legend}
            \addlegendentry{DSEDGITR}
            \addplot[fill=oiSkyBlue] table[x=graph,y=starplat_red]{Experiments/comparison_60_sssp/comp60.txt};
            \addplot[fill=oiBluishGreen] table[x=graph,y=starplat]{Experiments/comparison_60_sssp/comp60.txt};
            \addplot[fill=oiVermillion] table[x=graph,y=drone]{Experiments/comparison_60_sssp/comp60.txt};
            \addplot[fill=oiOrange] table[x=graph,y=galois]{Experiments/comparison_60_sssp/comp60.txt};
        \end{axis}
    \end{tikzpicture}
    \caption{SSSP: 60 procs, Setup 1}
\end{subfigure}
\hfill
\begin{subfigure}[t]{0.20\textwidth}
    \begin{tikzpicture}
        \begin{axis}[symbolic x coords={TW, OK, WK, LJ, PK, RM, UR},xtick=data, height=5.5cm,
            x tick label style={
        rotate=45,
        anchor=east
    }, ybar=0pt, ymax=60, height=4.0cm, bar width=0.1cm, ylabel=time]
            \addplot[fill=oiBluishGreen] table[x=graph,y=starplat]{Experiments/WCC_60/comp60.txt};
            \addplot[fill=oiSkyBlue] table[x=graph,y=starplat_fused]{Experiments/WCC_60/comp60.txt};
            \addplot[fill=oiVermillion] table[x=graph,y=DRONE]{Experiments/WCC_60/comp60.txt};
            \addplot[fill=oiOrange] table[x=graph,y=galois]{Experiments/WCC_60/comp60.txt};
        \end{axis}
    \end{tikzpicture}
    \caption{WCC: 60 procs,Setup 1}
    \label{comp:wcc1}
\end{subfigure}
\hfill
\begin{subfigure}[t]{0.20\textwidth}
    \begin{tikzpicture}
        \begin{axis}[symbolic x coords={TW, OK, WK, LJ, PK, RM, UR},xtick=data, height=5.5cm,
            x tick label style={
        rotate=45,
        anchor=east
    }, ybar=0pt, ymax=60, height=4.0cm, bar width=0.1cm, ylabel=time]
            \addplot[fill=oiBluishGreen] table[x=graph,y=DSEDG]{Experiments/WCC_40/comp40.txt};
            \addplot[fill=oiBlue] table[x=graph,y=DSEDGITR]{Experiments/WCC_40/comp40.txt};
            \addplot[fill=oiVermillion] table[x=graph,y=DRN]{Experiments/WCC_40/comp40.txt};
            \addplot[fill=oiOrange] table[x=graph,y=Gal]{Experiments/WCC_40/comp40.txt};
        \end{axis}
    \end{tikzpicture}
    \caption{WCC: 40 procs,Setup 2}
    \label{comp:wcc2}
\end{subfigure}
\hfill
\begin{subfigure}[t]{0.20\textwidth}
    \begin{tikzpicture}
        \begin{axis}[symbolic x coords={ OK, WK, LJ, PK, RM, UR},xtick=data, height=5.5cm,
            x tick label style={
        rotate=45,
        anchor=east
    }, ybar=0pt, ymax=60, height=4.0cm, bar width=0.1cm, ylabel=time]
            \addplot[fill=oiBluishGreen] table[x=graph,y=DSEDG]{Experiments/SSSP_40/comp40.txt};
            \addplot[fill=oiSkyBlue] table[x=graph,y=DSEDGPLS]{Experiments/SSSP_40/comp40.txt};
            \addplot[fill=oiOrange] table[x=graph,y=GALpush]{Experiments/SSSP_40/comp40.txt};
            \addplot[fill=oiVermillion] table[x=graph,y=DRN]{Experiments/SSSP_40/comp40.txt};
        \end{axis}
    \end{tikzpicture}
    \caption{SSSP: 40 procs, Setup 2}
    \label{comp:sssp2}
\end{subfigure}
\ref{sharedlegend3}
\caption{Comparison of PR, SSSP, and WCC with prior art on social media networks.}
\textit{Note: DSEDG = StarDist; DSEDGPLS = StarDist with pulse optimization, DSEDGITR = StarDist with iterBFS.}
\label{comp:all}
\end{figure}

DRONE kept timing out in its graph partitioning phase when operated at our maximum compute capacity (90 processes) on our 3 node setup. Therefore, we run two sets of comparison. One compares StarDist-generated code with and without pulse fusion only with Galois and the other compares StarDist generated code with both Galois and DRONE on our second setup (Table~\ref{tab:comp}). We provide a comparison between Galois, DRONE, and StarDist in Figure~\ref{comp:all} for 60 processes spread across two nodes. However, two nodes is a small cluster for demonstrating our generated code. Therefore, in Table~\ref{tab:comp}, we provide results of only DRONE's evaluation on our second setup. 
Both Galois and DRONE use hardware acceleration. Galois uses hidden \texttt{pthreads} to communicate updates and maintain their graph structure, and DRONE's repository has a GPU-accelerated version using NCCL and CUDA (which is not mentioned in the publication). 
We found that DRONE's SSSP repository pushed every vertex divisible by 100 into the initial reduction queue, leading to incorrect shortest paths, but faster convergence. We corrected the same by altering \href{https://github.com/ICT-ACL/DRONE/blob/d43afc42d9ac45e1b948a86167c15dc802a4e1cb/DRONE/src/algorithm/SSSP.go\#L66}{line 66} in the algorithm header of their repository to push only the source vertex into the priority queue during initialization. We also could not get DRONE's PR algorithm to work on graphs with dangling vertices (vertices with outdegree 0). This situation occurs in all our graphs. So, we could not produce a comparison with DRONE, and present PR results with Galois alone.

Our work focuses on optimizing the communication pattern. Leveraging hardware acceleration using \texttt{OpenMP} or \texttt{CUDA} within the \texttt{OpenMPI} framework for intra-node activities is feasible but obscures away communication patterns that require optimization. StarPlat had a scaling degeneration following the introduction of RMA structures for backing graph properties and therefore, the latest version could not be used for comparison in this context.
To this end, we alter the hidden \texttt{pthreads} in Galois used for intra-node graph traversal. To our surprise, we noted an improvement in performance of d-Galois in SSSP when multi-threading was switched off.
For comparing connected components, our fastest implementation has been through the \texttt{iterateBFS} construct that traverses the graph in level order. This construct, however, utilizes a distributed BFS tree. The algorithm iterates level order and takes updates through our reduction queue. We feel knowing the topology of the graph beforehand would not be a fair comparison and therefore provide results on StarDist code utilize simple vertex and neighbor traversals to calculate the weakly connected components as well.
StarDist PR performed better than Galois on all graphs. However, we note that Galois, to minimize communication, checks whether any vertex was modified in the previous step. For StarDist, we have utilized the standard early stopping condition. For real world power law graphs, the convergence using the standard stopping condition is better.

\begin{figure*}[t]
\begin{tabular}{|l|l|rrrrrrrrrr|r|}
\hline
Algo. & Variant & PK & OK & TW & SW & WK & GM & US & LJ & RM & UR & \textbf{AVG}\\
\hline
PR   & Gal(1) &  33.381 & 75.345 & 113.471 & 241.334 & 45.083 & 18.437 & 42.159 & 8.017 & 107.807 & 39.515 & 72.445\\
     & DEG (1) & 0.533 & 2.549 & 34.382 & 119.439 & 2.288 & 0.113873 & 0.298465 & 1.667 & 8.791 & 1.001 & \textbf{17.106}\\
\hline
SSSP & Gal (1) & 20.146 & 37.086 & 4.579 & 5.947 & 37.292 & 64.461  & 2357.185 & 25.040 & 16.025 & 35.351 & 260.311\\
     & SPLT (1)     & 15     & 104    & 152   & 0.4   & 35     & 27      & 2302     & 47     & 70     & 17     & 276.940\\  
     & DEG(1) & 4.381 & 32.805 & 0.087 & 0.064 & 10.366 & 29.256 & 357.425 & 12.411 & 16.003 & 11.336 & \textbf{47.413}\\
     & DEGP (1)& 4.465 & 38.117 & 0.078 & 0.093 & 14.612 & 10.966 & 1645.312 & 13.722 & 21.648 & 12.058 & 176.107 \\
     \cline{2-13}
     & DRN(2) &  22.012 & 84.267 & 0.150 & TLE & 49.296 & 1563.665 & TLE & 49.767 & TLE & 119.917 & 688.907\\
     & Gal(2) &  7.482 & 19.177 & 2.451 & 1.424 & 30.076 & 48.619 & TLE & 11.774 & 7.297 & 19.857 & 174.816\\
     & DEG(2) & 4.762 & 44.091 & 0.025 & 0.026 & 14.100 & 14.624 & 187.401 & 13.958 & 40.493 & 8.053 & \textbf{32.753}\\
     & DEGP(2) & 5.406 & 57.512 & 0.0509 & 0.0415 & 21.631 & 8.575 & TLE & 20.101 & 50.448 & 7.653 & 177.142\\
\hline
WCC  & Gal(1) & 9.237 & 11.729 & 7.326 & 23.242 & 10.493 & 70.971 & 6.728 & 11.772 & 13.267 & 21.603 & \textbf{18.637}\\
     & DEG(1) & 2.510 & 15.950 & TLE & TLE & 8.541 & 33.247 & 142.275 & 6.534 & 26.051 & 12.355 & 344.746\\
     & DEGP(1)&  5.342 & 19.775 & 38.612 & 339.759 & 12.000 & 13.410 & 135.021 & 12.922 & 38.596 & 2.704 & 61.814 \\
     & DEGI (1)& 1.692 & 4.754 & 34.435 & 148.187 &3.737 & 29.093 & 52.696 & 3.853 & 12.232 & 2.651 & 29.333\\
     \cline{2-13}
     & Gal(2) & 3.382 & 5.791 & 10.006 & 17.279 & 6.317 & 8.324 & 2.238 & 5.111 & 6.316 & 21.622 & \textbf{8.639}\\
     & DRN(2) & 4.668 & 17.614 & 7.325 & 23.090 & 11.695 & 731.008 & 1726.607 & 9.023 & 2.947 & 103.568 & 263.755\\
     & DEG(2) & 3.070 & 24.739 & 29.946 & 7.289 & 8.381 & 16.749 & 159.389 & 6.849 & 20.374 & 14.642 & 29.143\\
     & DEGP(2)& 3.349 & 37.006 & 15.051 & 4.649 & 7.714 & 18.786 & 162.695 & 7.157 & 20.939 & 16.673 & 29.402\\
     & DEGI (2)& 2.264 & 11.519 & 6.271 & 40.775 & 5.150 & 16.105 & 8.105 & 28.324 & 5.769 & 27.041 & 15.132\\
\hline
\end{tabular}
\caption{Running times (second) on different algorithms and different frameworks}
\textit{Note : (1) denotes our first setup with 90 processes across 3 nodes and (2) denotes our second setup with 40 cores across 8 nodes. DEG = StarDist, DEGP = StarDist with pulse optimization, DEGI = StarDist with iterBFS construct; Gal = Galois, DRN = DRONE}
\label{tab:comp}
\end{figure*}

\subsection{Discussion}
StarDist without the pulse optimization ends up outperforming DRONE and Galois objectively. d-Galois utilizes point-to-point MPI directives for performing reduction operations and does not capitalize on operations that can tolerate out-of-order updates and DRONE encounters a bottleneck in master-worker communication when the worker resides on remote nodes. Both frameworks propagate updates in \textit{pulses}, which we observe are not graph invariant. Strict temporal segregating local and global updates can lead to graph traversals exploring sub-optimal paths within the local subgraph partition. 
The partitioning algorithms of the frameworks also play a huge role. \texttt{DRONE}'s \texttt{EBV} partitioner results in partitions that are heavily skewed. While communication is a major bottleneck in distributed algorithms, our strong scaling results confirm that graph algorithms are highly dependent on intra-node activity as well.
StarDist's \textit{pulse} optimization massively improved scalability for Germany road network, but degraded performance for the US road network when compared to StarDist with \texttt{short-circuit} reads and writes. However, it is still 1.4$\times$ better than StarPlat and Galois. Without \texttt{pulse} optimization, it is 6.60$\times$ better than StarPlat and Galois on our Setup 1 and is 8.35$\times$ better than DRONE on Setup 2. Hence, even though code generated by StarDist performs faster, the \texttt{pulse} optimization proved to be graph dependent and did not generalise well.
Upon analysis, we found the US to be the most challenging graph in our test suite. Being a single connected component, the updates \texttt{ping-pong} between processes and work distribution becomes uneven despite graph partitioning, which often makes it worse. We suggest a remediation strategy of latency hidden updates for monotonic reduction operations for mitigating this problem.

\section{Related Works}
Most of the prior works on distributed graph algorithm frameworks are library backed, requiring users to arrange a sequence of API calls to map out their algorithm. 
\textbf{Distributed Galois}\cite{ref3} provides a framework leveraging gluon-async~\cite{ref1} that can be used to write peer-to-peer distributed graph algorithms.
\textbf{Gemini}~\cite{ref4} framework adapts to runtime and provides an effective edge centric graph partitioning scheme. It facilitates writing graph algorithms both in shared and distributed memory situations. Much like gluon async in distributed Galois, updates are propagated iteratively until convergence.
\textbf{DRONE}~\cite{ref2} explores a subgraph boundary synchronization scheme over a vertex-cut based partitioning, a master-slave architecture where the master node responsible for synchronization and bookkeeping. DRONE~\cite{ref2} claims better performance over Gluon~\cite{ref1} and Gemini~\cite{ref4}.
In all three of the above frameworks~\cite{ref1, ref4, ref2}, the propagation of updates to mirror nodes and their synchronization happens in pulses, as defined in Goldberg and Tarjan's parallel implementation of push relabel~\cite{ref8}.
StarPlat~\cite{ref6} provides a unified high performant, domain specific, compiled language with an Abstract Syntax Tree representing semantics capturing graph entities and operations. However, it struggles with generating optimized code for distributed graph algorithms.
\textbf{Poly}~\cite{ref10} evaluated multiple possible reduction operations were to perform 2.21 $\times$ faster than the control. \citet{ref28} introduced a compiler analysis framework that relies on reordering of commutative operations to attain higher degree of parallelism in the generated code.
 Irregular memory access patterns with graph algorithms and their effects when scaled are studied in Cagra~\cite{ref9}.
\textbf{Diffuse}~\cite{ref17} uses constraints to fuse distributed tasks into chunks improving performance over \texttt{cuPy}~\cite{ref18} and has successfully incorporated their fusion logic into Legion's~\cite{legion} JIT compiler~\cite{ref19}.

There have been alternative strategies suggested to solve the IO bottleneck in sequential graph algorithms as well, FlashVM~\cite{ref29} provided high speed swap space architecture for improving workloads with irregular IO access patterns. Out-of-core solutions \cite{graphchi,Accooc} for graph algorithms operating on large graphs also achieve good execution times.


\section{Conclusion}
We built StarDist, an efficient code generator for distributed graph algorithms. Its analysis identifies reduction patterns and fuses them into pulses. Based on vertex residency, it also optimizes propagation of updates. Using a suite of graphs with varying characteristics, we illustrate that StarDist outperforms \texttt{DRONE} and \texttt{dGalois} with significant improvements. StarDist can be explored for complex graph algorithms.

\section{Acknowledgments}
Gemini 3.1 enterprise was used for simulating peer reviews. The authors thank Atharva Chougule and Rohan Kumar for initial inputs. Further advancements and contributions to this work are expected.


\bibliographystyle{ACM-Reference-Format}
\bibliography{software}

@inproceedings{ref1,
  author    = {Roshan Dathathri and Gurbinder Gill and Loc Hoang and Hoang-Vu Dang and Vishwesh Jatala and V. Krishna Nandivada and Marc Snir and Keshav Pingali},
  title     = {{Gluon-Async}: A Bulk-Asynchronous System for Distributed and Heterogeneous Graph Analytics},
  booktitle = {Proceedings of the 28th International Conference on Parallel Architectures and Compilation Techniques (PACT)},
  series    = {PACT '19},
  year      = {2019},
  pages     = {1--12},
  address   = {Seattle, WA, USA},
  publisher = {IEEE},
  doi       = {10.1109/PACT.2019.00010}
}

@inproceedings{ref2,
author = {Hjelm, Nathan},
title = {An Evaluation of the One-Sided Performance in Open MPI},
year = {2016},
isbn = {9781450342346},
publisher = {Association for Computing Machinery},
address = {New York, NY, USA},
url = {https://doi.org/10.1145/2966884.2966890},
doi = {10.1145/2966884.2966890},
booktitle = {Proceedings of the 23rd European MPI Users' Group Meeting},
pages = {184–187},
numpages = {4},
location = {Edinburgh, United Kingdom},
series = {EuroMPI '16}
}

@inproceedings{ref3,
author = {Nguyen, Donald and Lenharth, Andrew and Pingali, Keshav},
title = {A lightweight infrastructure for graph analytics},
year = {2013},
isbn = {9781450323888},
publisher = {Association for Computing Machinery},
address = {New York, NY, USA},
url = {https://doi.org/10.1145/2517349.2522739},
doi = {10.1145/2517349.2522739},
booktitle = {Proceedings of the Twenty-Fourth ACM Symposium on Operating Systems Principles},
pages = {456–471},
numpages = {16},
location = {Farminton, Pennsylvania},
series = {SOSP '13}
}

@inproceedings{ref4,
author = {Zhu, Xiaowei and Chen, Wenguang and Zheng, Weimin and Ma, Xiaosong},
title = {Gemini: a computation-centric distributed graph processing system},
year = {2016},
isbn = {9781931971331},
publisher = {USENIX Association},
address = {USA},
booktitle = {Proceedings of the 12th USENIX Conference on Operating Systems Design and Implementation},
pages = {301–316},
numpages = {16},
location = {Savannah, GA, USA},
series = {OSDI'16}
}

@article{ref5,
author = {Behera, Nibedita and Kumar, Ashwina and Rajadurai T, Ebenezer and Nitish, Sai and M, Rajesh Pandian and Nasre, Rupesh},
title = {{StarPlat: A versatile DSL for graph analytics}},
year = {2024},
issue_date = {Dec 2024},
publisher = {Academic Press, Inc.},
address = {USA},
volume = {194},
number = {C},
issn = {0743-7315},
url = {https://doi.org/10.1016/j.jpdc.2024.104967},
doi = {10.1016/j.jpdc.2024.104967},
journal = {J. Parallel Distrib. Comput.},
month = dec,
numpages = {17}
}

@misc{ref6,
      title={Generating Dynamic Graph Algorithms for Multiple Backends for a Graph DSL}, 
      author={Nibedita Behera and Ashwina Kumar and Atharva Chougule and Mohammed Shan P S and Rushabh Nirdosh Lalwani and Rupesh Nasre},
      year={2025},
      eprint={2507.11094},
      archivePrefix={arXiv},
      primaryClass={cs.DC},
      url={https://arxiv.org/abs/2507.11094}, 
}

@inproceedings{ref7,
author = {Graham, Richard L. and Woodall, Timothy S. and Squyres, Jeffrey M.},
title = {Open MPI: a flexible high performance MPI},
year = {2005},
isbn = {3540341412},
publisher = {Springer-Verlag},
address = {Berlin, Heidelberg},
url = {https://doi.org/10.1007/11752578\_29},
doi = {10.1007/11752578\_29},
booktitle = {Proceedings of the 6th International Conference on Parallel Processing and Applied Mathematics},
pages = {228–239},
numpages = {12},
location = {Pozna\'{n}, Poland},
series = {PPAM'05}
}

@article{ref8,
title = "A New Approach to the Maximum-Flow Problem",
author = "Goldberg, \{Andrew V.\} and Tarjan, \{Robert E.\}",
note = "Copyright: Copyright 2016 Elsevier B.V., All rights reserved.",
year = "1988",
month = oct,
day = "1",
doi = "10.1145/48014.61051",
language = "English (US)",
volume = "35",
pages = "921--940",
journal = "Journal of the ACM (JACM)",
issn = "0004-5411",
publisher = "Association for Computing Machinery (ACM)",
number = "4",
}

@INPROCEEDINGS {ref9,
author = { Zhang, Yunming and Kiriansky, Vladimir and Mendis, Charith and Amarasinghe, Saman and Zaharia, Matei },
booktitle = { 2017 IEEE International Conference on Big Data (Big Data) },
title = {{ Making caches work for graph analytics }},
year = {2017},
volume = {},
ISSN = {},
publisher = {IEEE},
address = {IEEE International Conference on Big Data},
pages = {293-302}
}

@misc{ref10,
      title={Polly's Polyhedral Scheduling in the Presence of Reductions}, 
      author={Johannes Doerfert and Kevin Streit and Sebastian Hack and Zino Benaissa},
      year={2015},
      eprint={1505.07716},
      archivePrefix={arXiv},
      primaryClass={cs.PL},
      url={https://arxiv.org/abs/1505.07716}, 
}

@inproceedings{ref11,
author = {Graham, Susan L. and Kessler, Peter B. and Mckusick, Marshall K.},
title = {Gprof: A call graph execution profiler},
year = {1982},
isbn = {0897910745},
publisher = {Association for Computing Machinery},
address = {New York, NY, USA},
url = {https://doi.org/10.1145/800230.806987},
doi = {10.1145/800230.806987},
booktitle = {Proceedings of the 1982 SIGPLAN Symposium on Compiler Construction},
pages = {120–126},
numpages = {7},
location = {Boston, Massachusetts, USA},
series = {SIGPLAN '82}
}

@article{ref12,
author = {Adhianto, L. and Banerjee, S. and Fagan, M. and Krentel, M. and Marin, G. and Mellor-Crummey, J. and Tallent, N. R.},
title = {HPCTOOLKIT: tools for performance analysis of optimized parallel programs http://hpctoolkit.org},
year = {2010},
issue_date = {April 2010},
publisher = {John Wiley and Sons Ltd.},
address = {GBR},
volume = {22},
number = {6},
issn = {1532-0626},
journal = {Concurr. Comput.: Pract. Exper.},
month = apr,
pages = {685–701},
numpages = {17}
}

@inproceedings{ref17,
author = {Yadav, Rohan and Sundram, Shiv and Lee, Wonchan and Garland, Michael and Bauer, Michael and Aiken, Alex and Kjolstad, Fredrik},
title = {Composing Distributed Computations Through Task and Kernel Fusion},
year = {2025},
isbn = {9798400706981},
publisher = {Association for Computing Machinery},
address = {New York, NY, USA},
url = {https://doi.org/10.1145/3669940.3707216},
doi = {10.1145/3669940.3707216},
booktitle = {Proceedings of the 30th ACM International Conference on Architectural Support for Programming Languages and Operating Systems, Volume 1},
pages = {182–197},
numpages = {16},
location = {Rotterdam, Netherlands},
series = {ASPLOS '25}
}

@inproceedings{ref18,
  author    = "Okuta, Ryosuke and Unno, Yuya and Nishino, Daisuke and Hido, Shohei and Loomis, Crissman",
  title     = "CuPy: A NumPy-Compatible Library for NVIDIA GPU Calculations",
  booktitle = "Proceedings of Workshop on Machine Learning Systems (LearningSys) in The Thirty-first Annual Conference on Neural Information Processing Systems (NIPS)",
  publisher = {Curran Associates Inc.},
  year      = "2017",
  url       = "http://learningsys.org/nips17/assets/papers/paper_16.pdf",
  address = {San Diego, USA}
}

@inproceedings{ref19,
author = {Bauer, Michael and Treichler, Sean and Slaughter, Elliott and Aiken, Alex},
title = {Legion: expressing locality and independence with logical regions},
year = {2012},
isbn = {9781467308045},
publisher = {IEEE Computer Society Press},
address = {Washington, DC, USA},
booktitle = {Proceedings of the International Conference on High Performance Computing, Networking, Storage and Analysis},
articleno = {66},
numpages = {11},
location = {Salt Lake City, Utah},
series = {SC '12}
}

@inproceedings{ref22,
author = {Munagala, Kameshwar and Ranade, Abhiram},
title = {I/O-complexity of graph algorithms},
year = {1999},
isbn = {0898714346},
publisher = {Society for Industrial and Applied Mathematics},
address = {USA},
booktitle = {Proceedings of the Tenth Annual ACM-SIAM Symposium on Discrete Algorithms},
pages = {687–694},
numpages = {8},
location = {Baltimore, Maryland, USA},
series = {SODA '99}
}

@inproceedings{ref24,
author = {Li, Feng and Das, Sudipto and Syamala, Manoj and Narasayya, Vivek R.},
title = {Accelerating Relational Databases by Leveraging Remote Memory and RDMA},
year = {2016},
isbn = {9781450335317},
publisher = {Association for Computing Machinery},
address = {New York, NY, USA},
url = {https://doi.org/10.1145/2882903.2882949},
doi = {10.1145/2882903.2882949},
booktitle = {Proceedings of the 2016 International Conference on Management of Data},
pages = {355–370},
numpages = {16},
location = {San Francisco, California, USA},
series = {SIGMOD '16}
}

@misc{ref26,
  title        = {Out Of Memory Handling},
  author       = {{Linux Kernel Documentation}},
  howpublished = {\url{https://www.kernel.org/doc/html/latest/admin-guide/mm/oom_kill.html}},
  year         = {2024}
}

@article{ref28,
author = {Rinard, Martin C. and Diniz, Pedro C.},
title = {Commutativity analysis: a new analysis technique for parallelizing compilers},
year = {1997},
issue_date = {Nov. 1997},
publisher = {Association for Computing Machinery},
address = {New York, NY, USA},
volume = {19},
number = {6},
issn = {0164-0925},
url = {https://doi.org/10.1145/267959.269969},
doi = {10.1145/267959.269969},
journal = {ACM Trans. Program. Lang. Syst.},
month = nov,
pages = {942–991},
numpages = {50}
}

@inproceedings{ref29,
author = {Saxena, Mohit and Swift, Michael M.},
title = {FlashVM: virtual memory management on flash},
year = {2010},
publisher = {USENIX Association},
address = {USA},
booktitle = {Proceedings of the 2010 USENIX Conference on USENIX Annual Technical Conference},
pages = {14},
numpages = {1},
location = {Boston, MA},
series = {USENIXATC'10}
}

@inbook{ref30,
author = {Deepayan Chakrabarti and Yiping Zhan and Christos Faloutsos},
title = {R-MAT: A Recursive Model for Graph Mining},
booktitle = {Proceedings of the 2004 SIAM International Conference on Data Mining (SDM)},
chapter = {},
pages = {442-446},
year = {2004},
publisher = {Proceedings of the 2004 SIAM International Conference on Data Mining},
address={Florida, USA},
doi = {10.1137/1.9781611972740.43},
URL = {https://epubs.siam.org/doi/abs/10.1137/1.9781611972740.43},
eprint = {https://epubs.siam.org/doi/pdf/10.1137/1.9781611972740.43}
}

@book{ref31,
  author    = {Jeremy G. Siek and Lie-Quan Lee and Andrew Lumsdaine},
  title     = {The Boost Graph Library: User Guide and Reference Manual},
  year      = {2002},
  publisher = {Addison-Wesley},
  address   = {Boston, MA}
}

@inproceedings{ref32,
author = {Riley, George F. and Jaafar, Talal M. and Fujimoto, Richard M. and Ammar, Mostafa H.},
title = {Space-parallel network simulations using ghosts},
year = {2004},
isbn = {0769521118},
publisher = {Association for Computing Machinery},
address = {New York, NY, USA},
url = {https://doi.org/10.1145/1013329.1013357},
doi = {10.1145/1013329.1013357},
booktitle = {Proceedings of the Eighteenth Workshop on Parallel and Distributed Simulation},
pages = {170–177},
numpages = {8},
location = {Kufstein, Austria},
series = {PADS '04}
}

@inproceedings{fregel,
author = {Emoto, Kento and Matsuzaki, Kiminori and Hu, Zhenjiang and Morihata, Akimasa and Iwasaki, Hideya},
title = {Think like a vertex, behave like a function! a functional DSL for vertex-centric big graph processing},
year = {2016},
isbn = {9781450342193},
publisher = {Association for Computing Machinery},
address = {New York, NY, USA},
url = {https://doi.org/10.1145/2951913.2951938},
doi = {10.1145/2951913.2951938},
booktitle = {Proceedings of the 21st ACM SIGPLAN International Conference on Functional Programming},
pages = {200–213},
numpages = {14},
location = {Nara, Japan},
series = {ICFP 2016}
}

@inproceedings{Pregel,
author = {Malewicz, Grzegorz and Austern, Matthew H. and Bik, Aart J.C and Dehnert, James C. and Horn, Ilan and Leiser, Naty and Czajkowski, Grzegorz},
title = {Pregel: a system for large-scale graph processing},
year = {2010},
isbn = {9781450300322},
publisher = {Association for Computing Machinery},
address = {New York, NY, USA},
url = {https://doi.org/10.1145/1807167.1807184},
doi = {10.1145/1807167.1807184},
booktitle = {Proceedings of the 2010 ACM SIGMOD International Conference on Management of Data},
pages = {135–146},
numpages = {12},
location = {Indianapolis, Indiana, USA},
series = {SIGMOD '10}
}

@article{DRONE,
  author       = {Shuai Zhang and
                  Zite Jiang and
                  Xingzhong Hou and
                  Mingyu Li and
                  Mengting Yuan and
                  Haihang You},
  title        = {{DRONE:} An Efficient Distributed Subgraph-Centric Framework for Processing
                  Large-Scale Power-law Graphs},
  journal      = {{IEEE} Trans. Parallel Distributed Syst.},
  volume       = {34},
  number       = {2},
  pages        = {463--474},
  year         = {2023},
  url          = {https://doi.org/10.1109/TPDS.2022.3223068},
  doi          = {10.1109/TPDS.2022.3223068},
  bibsource    = {dblp computer science bibliography, https://dblp.org}
}

@article{snap,
author = {Leskovec, Jure and Sosi\v{c}, Rok},
title = {SNAP: A General-Purpose Network Analysis and Graph-Mining Library},
year = {2016},
issue_date = {January 2017},
publisher = {Association for Computing Machinery},
address = {New York, NY, USA},
volume = {8},
number = {1},
issn = {2157-6904},
url = {https://doi.org/10.1145/2898361},
doi = {10.1145/2898361},
journal = {ACM Trans. Intell. Syst. Technol.},
month = jul,
articleno = {1},
numpages = {20}
}

@inproceedings{RMA,
author = {Li, M. and Potluri, S. and Hamidouche, K. and Jose, J. and Panda, D. K.},
title = {Efficient and truly passive MPI-3 RMA using InfiniBand atomics},
year = {2013},
isbn = {9781450319034},
publisher = {Association for Computing Machinery},
address = {New York, NY, USA},
url = {https://doi.org/10.1145/2488551.2488573},
doi = {10.1145/2488551.2488573},
booktitle = {Proceedings of the 20th European MPI Users' Group Meeting},
pages = {91–96},
numpages = {6},
location = {Madrid, Spain},
series = {EuroMPI '13}
}

@book{CONGEST,
  author    = {David Peleg},
  title     = {Distributed Computing: A Locality-Sensitive Approach},
  publisher = {SIAM},
  year      = {2000},
  doi       = {10.1137/1.9780898719772}
}

@article{mapred,
author = {Dean, Jeffrey and Ghemawat, Sanjay},
title = {MapReduce: simplified data processing on large clusters},
year = {2008},
issue_date = {January 2008},
publisher = {Association for Computing Machinery},
address = {New York, NY, USA},
volume = {51},
number = {1},
issn = {0001-0782},
url = {https://doi.org/10.1145/1327452.1327492},
doi = {10.1145/1327452.1327492},
journal = {Commun. ACM},
month = jan,
pages = {107–113},
numpages = {7}
}

@article{scalegraph,
author = {Sahu, Siddhartha and Mhedhbi, Amine and Salihoglu, Semih and Lin, Jimmy and \"{O}zsu, M. Tamer},
title = {The ubiquity of large graphs and surprising challenges of graph processing},
year = {2017},
issue_date = {December 2017},
publisher = {VLDB Endowment},
volume = {11},
number = {4},
issn = {2150-8097},
url = {https://doi.org/10.1145/3186728.3164139},
doi = {10.1145/3186728.3164139},
journal = {Proc. VLDB Endow.},
month = dec,
pages = {420–431},
numpages = {12}
}

@inproceedings{GreenMarl,
author = {Hong, Sungpack and Chafi, Hassan and Sedlar, Edic and Olukotun, Kunle},
title = {Green-Marl: a DSL for easy and efficient graph analysis},
year = {2012},
isbn = {9781450307598},
publisher = {Association for Computing Machinery},
address = {New York, NY, USA},
url = {https://doi.org/10.1145/2150976.2151013},
doi = {10.1145/2150976.2151013},
booktitle = {Proceedings of the Seventeenth International Conference on Architectural Support for Programming Languages and Operating Systems},
pages = {349–362},
numpages = {14},
location = {London, England, UK},
series = {ASPLOS XVII}
}

@inproceedings{Galois,
author = {Nguyen, Donald and Lenharth, Andrew and Pingali, Keshav},
title = {A lightweight infrastructure for graph analytics},
year = {2013},
isbn = {9781450323888},
publisher = {Association for Computing Machinery},
address = {New York, NY, USA},
url = {https://doi.org/10.1145/2517349.2522739},
doi = {10.1145/2517349.2522739},
booktitle = {Proceedings of the Twenty-Fourth ACM Symposium on Operating Systems Principles},
pages = {456–471},
numpages = {16},
location = {Farminton, Pennsylvania},
series = {SOSP '13}
}

@inproceedings{Galois2,
author = {Pingali, Keshav},
title = {High-speed graph analytics with the galois system},
year = {2014},
isbn = {9781450326544},
publisher = {Association for Computing Machinery},
address = {New York, NY, USA},
url = {https://doi.org/10.1145/2567634.2567648},
doi = {10.1145/2567634.2567648},
booktitle = {Proceedings of the First Workshop on Parallel Programming for Analytics Applications},
pages = {41–42},
numpages = {2},
location = {Orlando, Florida, USA},
series = {PPAA '14}
}

@inproceedings{Ligra,
author = {Shun, Julian and Blelloch, Guy E.},
title = {Ligra: a lightweight graph processing framework for shared memory},
year = {2013},
isbn = {9781450319225},
publisher = {Association for Computing Machinery},
address = {New York, NY, USA},
url = {https://doi.org/10.1145/2442516.2442530},
doi = {10.1145/2442516.2442530},
booktitle = {Proceedings of the 18th ACM SIGPLAN Symposium on Principles and Practice of Parallel Programming},
pages = {135–146},
numpages = {12},
location = {Shenzhen, China},
series = {PPoPP '13}
}

@article{survey,
author = {Meng, Lingkai and Shao, Yu and Yuan, Long and Lai, Longbin and Cheng, Peng and Li, Xue and Yu, Wenyuan and Zhang, Wenjie and Lin, Xuemin and Zhou, Jingren},
title = {A Survey of Distributed Graph Algorithms on Massive Graphs},
year = {2024},
issue_date = {February 2025},
publisher = {Association for Computing Machinery},
address = {New York, NY, USA},
volume = {57},
number = {2},
issn = {0360-0300},
url = {https://doi.org/10.1145/3694966},
doi = {10.1145/3694966},
journal = {ACM Comput. Surv.},
month = oct,
articleno = {27},
numpages = {39}
}

@inproceedings{Faloutsos1999,
author = {Faloutsos, Michalis and Faloutsos, Petros and Faloutsos, Christos},
title = {On power-law relationships of the Internet topology},
year = {1999},
isbn = {1581131356},
publisher = {Association for Computing Machinery},
address = {New York, NY, USA},
url = {https://doi.org/10.1145/316188.316229},
doi = {10.1145/316188.316229},
booktitle = {Proceedings of the Conference on Applications, Technologies, Architectures, and Protocols for Computer Communication},
pages = {251–262},
numpages = {12},
location = {Cambridge, Massachusetts, USA},
series = {SIGCOMM '99}
}

@inproceedings{powergraph,
author = {Gonzalez, Joseph E. and Low, Yucheng and Gu, Haijie and Bickson, Danny and Guestrin, Carlos},
title = {PowerGraph: distributed graph-parallel computation on natural graphs},
year = {2012},
isbn = {9781931971966},
publisher = {USENIX Association},
address = {USA},
booktitle = {Proceedings of the 10th USENIX Conference on Operating Systems Design and Implementation},
pages = {17–30},
numpages = {14},
location = {Hollywood, CA, USA},
series = {OSDI'12}
}

@article{mfa,
  title = {Monotone Data Flow Analysis Frameworks},
  volume = {7},
  year = {1977},
  author = {John B. Kam and Jeffrey D. Ullman},
  journal = {Acta Informatica},
  pages = {305--317},
}

@inproceedings{legion,
author = {Bauer, Michael and Treichler, Sean and Slaughter, Elliott and Aiken, Alex},
title = {Legion: expressing locality and independence with logical regions},
year = {2012},
isbn = {9781467308045},
publisher = {IEEE Computer Society Press},
address = {Washington, DC, USA},
booktitle = {Proceedings of the International Conference on High Performance Computing, Networking, Storage and Analysis},
articleno = {66},
numpages = {11},
location = {Salt Lake City, Utah},
series = {SC '12}
}

@inproceedings {graphchi,
author = {Aapo Kyrola and Guy Blelloch and Carlos Guestrin},
title = {{GraphChi}: {Large-Scale} Graph Computation on Just a {PC}},
booktitle = {10th USENIX Symposium on Operating Systems Design and Implementation (OSDI 12)},
year = {2012},
isbn = {978-1-931971-96-6},
address = {Hollywood, CA},
pages = {31--46},
url = {https://www.usenix.org/conference/osdi12/technical-sessions/presentation/kyrola},
publisher = {USENIX Association},
month = oct
}

@article{Accooc,
author = {Liu, Huanghai and Wang, Qinggang and Li, Huize and Zheng, Long and Si, Liwei and Zhao, Xu and Liao, Xiaofei and Jin, Hai and Xue, Jingling},
title = {Accelerating Out-of-Core Graph Random Walk Processing via Locality-Aware Algorithm-Hardware Co-Design},
year = {2026},
issue_date = {March 2026},
publisher = {Association for Computing Machinery},
address = {New York, NY, USA},
volume = {23},
number = {1},
issn = {1544-3566},
url = {https://doi.org/10.1145/3779123},
doi = {10.1145/3779123},
journal = {ACM Trans. Archit. Code Optim.},
month = mar,
articleno = {8},
numpages = {26}
}




\end{document}